\documentclass[prx,reprint,amsmath,amssymb,longbibliography]{revtex4-1}

\usepackage{graphicx}
\usepackage{dcolumn}
\usepackage{bm}
\usepackage{float}
\usepackage{color}
\usepackage{refstyle}
\usepackage{mathrsfs}
\usepackage{esint}
\usepackage[unicode=true,pdfusetitle,bookmarks=true,
bookmarksnumbered=false,bookmarksopen=false,breaklinks=false,
pdfborder={0 0 0},backref=false,colorlinks=true,citecolor=red]
{hyperref}
\usepackage{cleveref}
\providecommand\eqref[1]{\ref{eq:#1}}
\renewcommand\b[1]{{\bf  #1}}
\renewcommand\vec[1]{\boldsymbol{#1}}
\renewcommand\phi{\varphi}
\newcommand\tr{\mathrm{tr}}
\newcommand\del{\nabla}
\newcommand\e{\epsilon}

\newcommand\dd{\mathrm{d}}
\allowdisplaybreaks[3]

\begin{document}

\title{Hydrodynamics of Active Defects: from order to chaos to defect ordering}
\author{Suraj Shankar$^1$ and M.~Cristina Marchetti$^2$}
\affiliation{
$^1$Department of Physics, Harvard University, Cambridge, MA 02318, USA\\
$^2$Department of Physics, University of California Santa Barbara, Santa Barbara, CA 93106}

\date{\today}

\begin{abstract}
	Topological defects play a prominent role in the physics of two-dimensional materials. When driven out of equilibrium in active nematics, disclinations can acquire spontaneous self-propulsion and drive self-sustained flows upon proliferation. Here we construct a general hydrodynamic theory for a two-dimensional active nematic interrupted by a large number of such defects. Our equations describe the flows and spatio-temporal defect chaos characterizing active turbulence, even close to the defect unbinding transition. At high activity, nonequilibrium torques combined with many-body screening cause the active disclinations to spontaneously break rotational symmetry forming a collectively moving defect ordered polar liquid. By recognizing defects as the relevant quasiparticle excitations, we construct a comprehensive phase diagram for two-dimensional active nematics. Using our hydrodynamic approach, we additionally show that activity gradients can act like ``electric fields'', driving the sorting of topological charge. This demonstrates the versatility of our continuum model and its relevance for quantifying the use of spatially inhomogeneous activity for controlling active flows and for the fabrication of active devices with targeted transport capabilities.
\end{abstract}

\maketitle
\section{Introduction}
In recent years, the framework of active fluids has had significant success in describing the emergent collective motion of bacterial and cellular assemblies \cite{wensink2012meso,saw2017topological,kawaguchi2017topological,blanch2018turbulent,li2019data}. Such spontaneous self-organization into complex patterns on large scales is a characteristic feature of active matter, whose constituent units are individually self-driven \cite{ramaswamy2010mechanics,marchetti2013hydrodynamics}. Much interest has recently focused on active nematics \cite{ramaswamy2003active,doostmohammadi2018active} - collections of head-tail symmetric elongated units that exert active forces on their surroundings and organize in states with apolar orientational order. A rapidly growing list of experimental realizations of active nematics ranges from suspensions of cytoskeletal filaments and associated motor proteins \cite{sanchez2012spontaneous,keber2014topology,ellis2018curvature,kumar2018tunable} to vibrated granular rods \cite{narayan2007long}, and colonies of living cells \cite{Zhou2014,nishiguchi2017long,duclos2017topological,saw2017topological,kawaguchi2017topological,blanch2018turbulent,li2019data}. In all these systems the interplay of orientational order and self-sustained active flows yields a rich collection of dynamical states, including spontaneous laminar flows \cite{voituriez2005spontaneous,duclos2018spontaneous} and spatio-temporal chaos or ``active turbulence'' with the proliferation of topological defects \cite{thampi2013velocity,giomi2013defect,shankar2018defect}.

While a great deal of understanding has been gained from extensive numerical work \cite{thampi2013velocity,giomi2013defect,shi2013topological,thampi2014instabilities,gao2015multiscale,giomi2015geometry,hemingway2016correlation,alert2019universal} and experiments \cite{guillamat2017taming,lemma2018statistical,martinez2019selection}, the theoretical analysis of low Reynolds number turbulence in active nematics, beyond deterministic linear instabilities, remains an open problem. A notable exception is Ref.~\cite{giomi2015geometry}, where a phenomenological mean field theory was proposed to describe active nematic turbulence on short scales by using the fact that topological defects actively drive flow. In contrast to equilibrium nematic liquid crystals, disclinations of strength $\pm1/2$ are spontaneously generated in pairs by activity. Importantly, the comet-shaped $+1/2$ defect acquires self-propulsion \cite{giomi2013defect} which allows for an activity-driven defect unbinding transition to a turbulent state of spatio-temporal chaos \cite{shankar2018defect}. Studies of active suspensions of microtubule bundles have also reported a remarkable state where $+1/2$ defects may themselves orientationally order in a nematic fashion on length and time scales larger than the mean free path or lifetime of an individual defect \cite{decamp2015orientational}. This observation remains the subject of debate in the literature, as numerical simulations of continuum nematodynamic equations and of particle-based models predominantly report only \emph{polar} defect ordering \cite{decamp2015orientational,putzig2016instabilities,srivastava2016negative,patelli2019understanding} or defect lattices \cite{doostmohammadi2016stabilization}, whereas nematic defect order is found to be generally transient and short-lived \cite{oza2016antipolar,srivastava2016negative}. The absence of a clear physical picture for the mechanism of defect ordering has hindered a resolution of this debate. Finally, recent experiments have suggested that topological defects may serve a biological function as centers of cell extrusion or accumulation in epithelia \cite{saw2017topological,kawaguchi2017topological}, seed mound formation in bacteria \cite{yaman2018emergence}, or control the morphology of the interface of growing cell colonies~\cite{doostmohammadi2016defect}. The nature and dynamics of active defects have therefore been the subject of intense research efforts in recent times.

In this paper we focus on topological defects and formulate a theory of two dimensional (2D) active nematics on a substrate in terms of the large scale dynamics of an interacting gas of unbound disclinations. Our work yields a complete analytical phase diagram that includes both the defect mediated melting of the active nematic and the defect ordering phase transition as a function of activity and noise. We additionally demonstrate that the defect hydrodynamic equations derived here provide a versatile framework for quantifying the behavior of systems with spatially varying activity.

In the remainder of the Introduction we motivate our approach and summarize the main results of our work.

\subsection{Dynamics of Active Nematics}
A now well accepted continuum description of active nematics builds on the established hydrodynamics of passive liquid crystals, augmented by nonequilibrium active stresses. This approach typically involves hydrodynamic equations for the fluid flow velocity $\b{u}$ and the nematic order parameter $\b{Q}$, a rank-2 symmetric traceless tensor in 2D, whose independent components are conveniently written in terms of a complex field
\begin{equation}
	\Psi(\b{r})=S(\b{r})e^{2i\theta(\b{r})}\ ,\label{eq:Psi}
\end{equation}
with $Q_{xx}=-Q_{yy}=S\cos2\theta$ and $Q_{xy}=Q_{yx}=S\sin2\theta$. Here $S$ is the scalar order parameter and $\theta$ is the angle of the nematic director $\hat{\b{n}}=(\cos\theta,\sin\theta)$.
\begin{figure}[]
	\centering
	\includegraphics[width=0.5\textwidth]{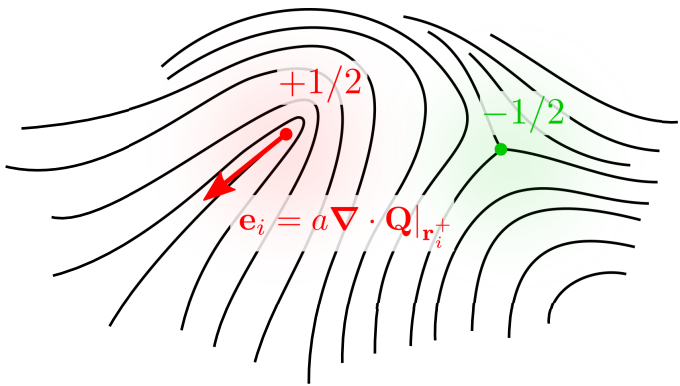}
	\caption{$+1/2$ (red) and $-1/2$ (green) disclinations in a 2D active nematic with the black lines following the local nematic order. The $+1/2$ defect has a local geometric polarity $\b{e}_i$ shown as a red arrow. Here  $a$ is the defect core size and the divergence is evaluated at the position $\b{r}_i^+$ of the defect core. }
	\label{fig:cartoon}
\end{figure}
Numerous numerical studies \cite{putzig2016instabilities,srivastava2016negative,hemingway2016correlation} have shown that this continuum model is capable of reproducing the phenomenology of 2D active nematics on a substrate, including the proliferation of topological defects, as both $\b{Q}$ and $\b{u}$ remain well defined even in the presence of defects. On the other hand, the order parameter field and flow velocity evolve in highly complex and nonlinear ways for large activity, impeding any analytical progress. As a consequence, a theory addressing both defect chaos and defect ordering has so far proved difficult.

An alternative and fruitful strategy has been to focus on the topological defects, which both in and out of equilibrium constitute elementary, yet nontrivial, excitations of the homogeneous ordered state. This approach explicitly recognizes defects as the relevant excitations driving complex active flows and aims at developing an effective description of defects as particles, akin to the well established mapping of point-like topological defects in 2D equilibrium systems onto a Coulomb gas \cite{chaikin2000principles}.
In 2D nematic liquid crystals the lowest energy topological defects are strength $\pm1/2$ point disclinations, corresponding to a distortion of the orientation where the angle $\theta$ becomes multivalued, acquiring a $\pm\pi$ jump as one encircles the respective defect, while the magnitude $S=1$ almost everywhere, vanishing only at the defect core.

The additional challenge in 2D \emph{active} nematics is that the $+1/2$ disclination becomes motile, with the local geometric polarity of the defect, defined as $\b{e}_i=a\bm\del\cdot\b{Q}(\b{r}_i^+)$ \cite{shankar2018defect} (see Fig.~\ref{fig:cartoon}), dictating its polarization. The $-1/2$ disclination is not self-propelled by virtue of its three-fold symmetry.
 As a result, an effective particle model for defects must incorporate the angular dynamics of the $+1/2$ defect polarization. Although the geometric polarity of the $+1/2$ disclination is obviously present even in passive nematics, there it remains a fast mode, rapidly relaxing on a short time scale. In contrast, when \emph{active}, the dynamical nature of the polarity allows it to be a slow mode with qualitatively new physics.

Recently, working perturbatively in activity, we and collaborators mapped the dynamics of active defects onto that of a mixture of motile ($+1/2$) and passive ($-1/2$) particles with Coulomb-like interaction forces and aligning torques \cite{shankar2018defect}, putting on firm ground previous purely phenomenological models \cite{giomi2013defect,pismen2013dynamics,keber2014topology}. Using this effective model, we were able to determine the critical activity separating the quasi-ordered two-dimensional nematic from the turbulent-like state of unbound defects.
This previous work demonstrates that, like in 2D equilibrium systems,  treating defects as quasiparticles affords a new description of 2D ordered active media dual to the more conventional one based on order parameter fields. Far from equilibrium, where analytical progress is often limited, such techniques are very valuable and can provide insight into complex phenomena that may be hard to rationalize otherwise. In the present paper, we have adopted this viewpoint to build a unified theory for the dynamical states of 2D active nematics by considering topological defects as the relevant degrees of freedom that drive large scale flows in the system.


\subsection{Results and Outline}

A key new result of our work is the derivation of a hydrodynamic theory for active defects that describes defect dynamics on length scales larger than the mean defect separation or the nematic correlation length $\xi$. This is obtained by systematically coarse-graining the effective particle model for the defects derived in Ref.~\cite{shankar2018defect}.
We then show that our hydrodynamic theory provides a complete theoretical description of defect organization in active nematics.

\paragraph{Defect Hydrodynamics.} The defect hydrodynamic equations are constructed in the spirit of previous classic approaches used to study the dynamic response of superfluid films \cite{ambegaokar1980dynamics}, planar magnets and rotating Helium \cite{volovik1980hydrodynamics}, flux liquids \cite{marchetti1990hydrodynamics} and the melting of 2D crystals \cite{zippelius1980dynamics}. They are formulated in terms of continuum fields, given by the number and current densities of the $+1/2$ and $-1/2$ defects defined as
\begin{align}
	n_{\pm}(\b{r},t)&=\left\langle\sum_i\delta(\b{r}-\b{r}_i^{\pm}(t))\right\rangle\ ,\label{eq:npm}\\
	\b{j}_{\pm}(\b{r},t)&=\left\langle\sum_i\dot{\b{r}}_i^{\pm}\delta(\b{r}-\b{r}_i^{\pm}(t))\right\rangle\ ,\label{eq:jpm}
\end{align}
where $\b{r}_i^{\pm}$ and  $\dot{\b{r}}_i^{\pm}$ are the position and velocity of the $i$th $\pm1/2$ defect, respectively, 
and the phase gradient or ``superfluid velocity'' \footnote{Our presentation closely follows Refs.~\cite{ambegaokar1980dynamics,zippelius1980dynamics}}
\begin{equation}
	\b{v}_n=\dfrac{1}{2|\Psi|^2}\mathrm{Im}\left(\Psi^*\bm\nabla\Psi\right)=\bm\nabla\theta\ .\label{eq:vs}
\end{equation}
Note that, unlike in superfluid He films, where the analogue of $\b{v}_n$ genuinely represents the flow velocity of the condensate, here $\b{v}_n$ is simply the gradient of the angle of the director. In this sense, it captures the distortion of the nematic and is perhaps closer in spirit to the phase gradient defined in other liquid crystalline phase, such as the hexatic \cite{zippelius1980dynamics}, relevant to the study of 2D melting at equilibrium.

Now, in the presence of defects, the phase $\theta$ of the order parameter field $\Psi$ is multivalued, while $\b{v}_n$ remains smooth and single-valued everywhere away from the defect cores, and provides a useful description of director deformations.
Defects are continuously created and annihilated, hence the total number density of defects $n=(n_++n_-)/2$, is not conserved. On the other hand, since defects are created and annihilated only in pairs, the topological charge density $\rho=(n_+-n_-)/2$ is always conserved and related to $\b{v}_n$ through the important topological constraint 
\begin{equation}
	\hat{\b{z}}\cdot\left(\bm\del\times\b{v}_n\right)=2\pi \rho\ ,\label{eq:topo}
\end{equation}
that allows an analogy with superfluid hydrodynamics \cite{ambegaokar1980dynamics} and electrostatics, through Gauss' law \cite{Jackson1975}.
An important distinguishing feature of the 2D active nematic is that that the geometric polarization of the $+1/2$ disclination provides a new dynamically relevant internal degree of freedom. This requires the addition of a new hydrodynamic field, the defect polarization density, defined as
\begin{equation}
	\b{p}(\b{r},t)=\left\langle\sum_i\b{e}_i(t)\delta(\b{r}-\b{r}_i^+(t))\right\rangle\ .\label{eq:p}
\end{equation}
The polarization directs the self-propulsion of the $+1/2$ disclinations through self-induced active backflows. In addition, it experiences active torques that reorient the defect in response to the elastic forces from other defects. Both these properties underlie much of the phenomena explored here.

The defect hydrodynamic equations are presented in Sec.~\ref{sec:hydro} and their derivation is shown in Appendix~\ref{app:calc}. 

\paragraph{Isotropic defect chaos.} Upon analyzing the steady states of the defect hydrodynamic equations and their stability (Secs.~\ref{sec:states} and~\ref{sec:stability}), we find two  stable  states with a finite number of unbound defect. The first is the spatio-temporally chaotic state referred to as ``active turbulence''  in the literature \cite{doostmohammadi2018active}. In this state the $+1/2$ defects, although motile, have no preferred direction of polarization. We refer to this state as ``isotropic  defect chaos'', where `isotropic' refers here to the fact that the gas of $+1/2$ defects has zero mean polarization ($\langle\b{p}\rangle=\b{0}$). Our equations validate previous phenomenological scaling hypotheses and provide a well founded theoretical description of defect chaos in the active turbulent regime by extending the mean-field approach of Ref.~\cite{giomi2015geometry} to large scales. Many-body screening crucially impacts the decay of the conserved defect charge density due to the presence of the topological constraint given by Eq.~\ref{eq:topo}, unique to defect hydrodynamics. As the decay rate is governed by the density $n$ of free disclinations, we find that the dominant length scale of correlations in nematic order or velocity are all controlled mainly by the mean defect spacing $\xi\sim1/\sqrt{n}$, in agreement with previous numerical results \cite{giomi2015geometry,hemingway2016correlation}.
\begin{figure}[]
	\centering
	\includegraphics[width=0.45\textwidth]{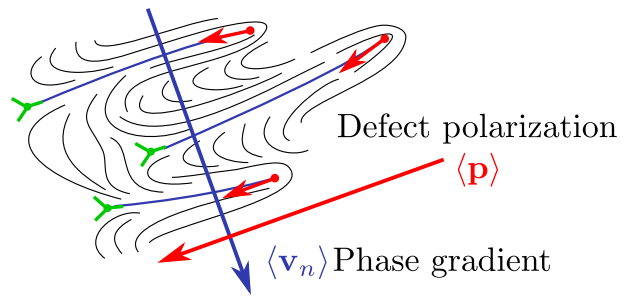}
	\caption{Structure of the polar defect ordered state for  extensile activity ($v<0$). The spontaneous polar ordering of the $+1/2$ defects combined with their active self-propulsion leads to a spontaneously flowing defect liquid. The nonvanishing average defect polarization ($\langle\b{p}\rangle\neq\b{0}$) simultaneously forces a nonvanishing phase gradient ($\langle\b{v}_n\rangle\neq\b{0}$) to condense in the orthogonal direction. This leads to a periodic array of kink walls (also called $\pi$- or N{\'e}el-walls) in the underlying nematic, with each kink wall (blue lines) terminating at the $+1/2$ (red) and $-1/2$ (green) defects.}
	\label{fig:flock}
\end{figure}

\begin{figure*}[]
	\centering
	\includegraphics[width=\textwidth]{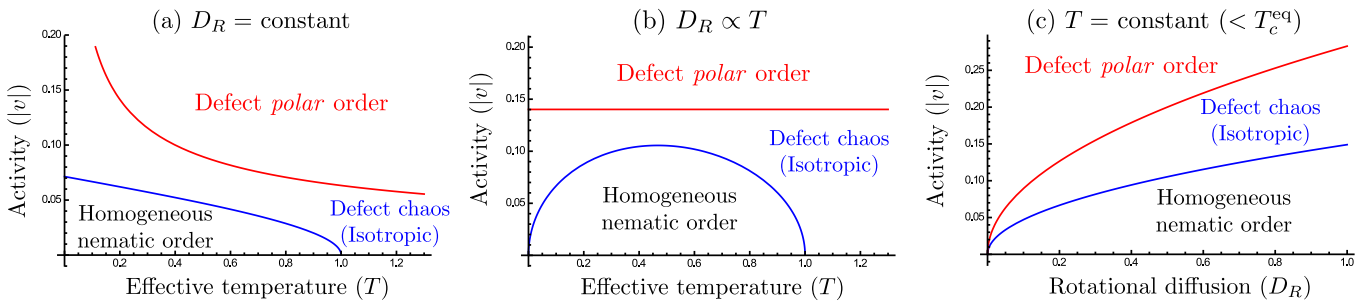}
	\caption{Full phase diagram as a function of activity ($|v|$), effective temperature ($T$) and rotational diffusion ($D_R$). For the more experimentally accessible phase diagram, with activity ($|v|$) versus liquid crystal elasticity ($K$), see Fig.~\ref{fig:phase_dgm_new}. We have fixed $\nu=\pi$ and $T_c^{\mathrm{eq}}=1$ in all the plots. (a) $|v|-T$ plane for fixed $D_R$, (b) $|v|-T$ plane for $D_R\propto T$ showing re-entrant melting as a function of $T$ and (c) $|v|-D_R$ plane for fixed $T$ below the equilibrium melting temperature ($T_c^{\mathrm{eq}}$). Two phase transition boundaries are marked in all three phase diagrams. The defect mediated melting transition (blue line) separates homogeneous nematic order from isotropic defect chaos. The defect ordering transition (red line) separates isotropic defect chaos (active turbulence) from the \emph{polar} defect ordered state consisting of a $+1/2$ defect flock. A cartoon of the structure of the polar defect flock is sketched in Fig.~\ref{fig:flock}.}
	\label{fig:phasedgm}
\end{figure*}

\paragraph{Polar defect order.} For higher activity, our equations predict a second stable state, where the motile $+1/2$ defects themselves order. This occurs via a continuous transition in which the active $+1/2$ disclinations collectively align and condense into a liquid with long-range \emph{polar} order. At the same time, the underlying nematic develops a periodic modulation of kink walls, as shown in the sketch of Fig.~\ref{fig:flock}. This state, which we refer to as ``polar defect order'', has \emph{no} giant fluctuations in either the defect charge or number density and provides an intriguing realization of a ``Malthusian defect flock''. While polar defect order has been reported before in numerical models of active nematics \cite{decamp2015orientational,putzig2016instabilities,srivastava2016negative,patelli2019understanding}, the mechanism driving it has remained unexplained. Our work identifies a mechanism for polar order as arising from both active self-aligning torques (derived perturbatively in activity in Ref.~\cite{shankar2018defect}) and many-body screening. Overall, our result is a complete analytical description of 2D active nematics that includes both the defect mediated melting of the active nematic and the defect ordering phase transition as a function of activity and noise. The various states and transitions are summarized in the phase diagram shown in Fig.~\ref{fig:phasedgm} in terms of parameters of the coarse-grained theory. To make contact with possible experiments, we have also reformulated the phase diagram in Fig.~\ref{fig:phase_dgm_new} in terms of activity and the liquid crystal stiffness, quantities that are both directly accessible and tunable in experiments. We note here that \emph{nematic} order of defects as reported in the experiments of Ref.~\cite{decamp2015orientational} cannot appear in our model via a transition from the isotropic defect chaos state. The reasons for this are discussed in Sec.~\ref{subsec:defectorder} (see also Fig.~\ref{fig:stability}) and further elaborated in Appendix~\ref{app:efield} using an analogy with electrostatics. In Sec.~\ref{sec:discussion}, we expand on future directions and current challenges, including the theoretical possibilities (or lack thereof) for the existence of apolar defect order.

\paragraph{Spatially-varying activity and defect trapping.} Finally, in Sec.~\ref{sec:inhomo} we demonstrate the versatility of our hydrodynamic approach by employing it to describe defect dynamics in systems with spatially inhomogeneous activity. A simple motif we study is an active-passive interface. Due to the self-propelled nature of the $+1/2$ disclinations, they are found to accumulate on the passive (low activity) side of the interface. The consequent charge segregation and local polarization at the interface indicates that an activity gradient can be thought of as a local ``electric field'' driving charge sorting. An extension of the same phenomenon is also realized at an extensile-contractile interface across which activity changes sign. Such basic principles can be combined with more complicated activity patterns to position and move defects in a programmable fashion. The use of activity gradients to control and guide defect dynamics is an important technique to develop active microfluidic devices with targeted transport capabilities. The spatio-temporal modulation of activity with light \cite{schuppler2016boundaries,ross2018controlling} is a very promising approach in functionalizing active matter to engineer new metamaterials. We expect our theoretical results and the proposed hydrodynamic framework to be useful tools in predicting the collective behavior of active defects in both such inhomogeneous backgrounds and complex geometries.

\section{Active defect hydrodynamics}
\label{sec:hydro}

We consider a 2D active nematic on a substrate with a finite concentration of unbound disclinations. Defects are unbound in pairs to maintain charge neutrality, hence the system contains an equal number of $+1/2$ and $-1/2$ defects.
In Ref.~\cite{shankar2018defect}, the full nematodynamic equations for an active nematic on a substrate were recast into an effective model for defects as interacting quasiparticles. While the particle model was explicitly derived only perturbatively for small activity, its general structure is expected to survive for large activity as well. Different models for active defect dynamics of varying complexity have been proposed by others as well \cite{cortese2018pair,tang2019theory} but the basic qualitative features remain the same. As the defect equations of motion presented in Ref.~\cite{shankar2018defect} are the easiest to interpret and most amenable to direct coarse-graining, we proceed with them to compute the required transport coefficients. The details of the calculation are given in Appendix~\ref{app:calc}. We shall present the final equations directly here.

The number of $+1/2$ or $-1/2$ defects can change through pair creation or annihilation events, hence the individual number densities $n_{\pm}$ evolve according to
\begin{equation}
	\partial_tn_{\pm}+\bm\del\cdot\b{j}_{\pm}=W_c-W_a\ ,\label{eq:npmeq}
\end{equation}
where $W_c$  and $W_a$ are the rates of defect creation and annihilation, respectively. Both $W_c$ and $W_a$ depend on activity and $n_{\pm}$; we refrain from specifying their explicit form which only serves as an input to the hydrodynamic theory.
The constitutive relation for the defect currents $\b{j}_{\pm}$ is (see Appendix~\ref{app:calc} for derivation)
\begin{align}
	\b{j}_+&=v\b{p}+\nu \kappa\,n_+\vec{\e}\cdot\b{v}_n-D_0\bm\del n_+\ ,\label{eq:jp}\\
	\b{j}_-&=-\nu \kappa\,n_-\vec{\e}\cdot\b{v}_n-D_0\bm\del n_-\ ,\label{eq:jm}
\end{align}
where $\vec{\e}$ is the 2D Levi-Civita tensor, $\nu=\pi\mu\gamma$ a dimensionless number involving the defect mobility $\mu$ and the rotational viscosity $\gamma$, and $D_0=\mu T$ is the bare defect diffusion constant ($T$ is the corresponding effective temperature). Liquid crystal elasticity controls the nematic diffusion constant $\kappa=K/\gamma$, where $K$ is a Frank elastic constant.
Activity is encoded in the self-propulsion speed $|v|$ of the $+1/2$ defect. Note that $v$ can be of either sign depending on the nature of active stresses in the medium. Extensile systems have $v<0$ and contractile systems have $v>0$. Even in the passive limit ($v=0$), defects move transverse to the local phase gradient $\b{v}_n$. This response is akin to the Magnus force on vortices \cite{lamb1993hydrodynamics} or the Peach-Koehler force on dislocations \cite{peach1950forces}.

Defect hydrodynamics crucially differs from conventional long-wavelength hydrodynamics due to the presence of a \emph{topological constraint}.
For an arbitrary closed curve $\Gamma$ enclosing $N$ unbound disclinations carrying charges $q_i=\pm1/2$ located at positions $\b{r}_i^{\pm}$, the net accumulated director phase is given by the line integral
\begin{equation}
	\oint_{\Gamma}\dd \theta\equiv\oint_{\Gamma}\dd\b{s}\cdot\b{v}_n=2\pi\sum_{i=1}^{N}q_i\ .
\end{equation}
Using Stokes' theorem and Eq.~\ref{eq:npm}, this gives the topological constraint given in Eq.~\ref{eq:topo}.
Using Eqs.~\ref{eq:npmeq},~\ref{eq:jp} and~\ref{eq:jm}, we obtain equations for the charge and number densities $\rho$ and $n$ as
\begin{gather}
	\partial_t\rho+\bm\del\cdot\b{j}_{\rho}=0\ ,\label{eq:rhoeq}\\
	\partial_tn+\bm\del\cdot\b{j}_n=W_c-W_a\ ,\label{eq:neq}\\
	\b{j}_{\rho}=\dfrac{(\b{j}_+-\b{j}_-)}{2}=\dfrac{v}{2}\b{p}+\nu\kappa\,n\,\vec{\e}\cdot\b{v}_n-D_0\bm\del\rho\ ,\label{eq:jrho}\\
	\b{j}_{n}=\dfrac{(\b{j}_++\b{j}_-)}{2}=\dfrac{v}{2}\b{p}+\nu\kappa\,\rho\,\vec{\e}\cdot\b{v}_n-D_0\bm\del n\ .\label{eq:jn}
\end{gather}
Finally, the dynamics of the phase gradient $\b{v}_n$ and of the defect polarization $\b{p}$ is derived by coarse-graining (see Appendix~\ref{app:calc}) to give
\begin{widetext}
\begin{gather}
	\partial_t\b{v}_n=2\pi\,\vec{\e}\cdot\b{j}_{\rho}+\kappa\bm\del(\bm\del\cdot\b{v}_n)\ ,\label{eq:vseq}\\
	\partial_t\b{p}-\nu\kappa\,\hat{\b{z}}\cdot(\b{v}_n\times\bm\del)\b{p}=-\left[D_R+2\pi\nu\kappa\,\rho+\beta|\b{v}_n|^2\right]\b{p}-v_2\,n_+\vec{\e}\cdot\b{v}_n-\dfrac{v_1}{2}\bm\del n_+-2\kappa(\del\cdot\b{v}_n)\vec{\e}\cdot\b{p}+D_0\del^2\b{p}\ .\label{eq:peq}
\end{gather}
\end{widetext}

In Eq.~\ref{eq:vseq}, the defect charge current $\b{j}_{\rho}$ explicitly breaks the ``conservation'' of the phase gradient $\b{v}_n$. This, as we shall see later, in turn also causes the charge density $\rho$ to relax on a \emph{finite} time-scale notwithstanding the local conservation of topological charge (Eq.~\ref{eq:rhoeq}). The second term on the right hand side of Eq.~\ref{eq:vseq} describes the relaxation of smooth director deformations due to liquid crystal elasticity.
For the polarization dynamics (Eq.~\ref{eq:peq}) we have adopted a simple isotropic closure for the second moment while neglecting all higher order correlations (see Appendix~\ref{app:calc}). This is sufficient for our purposes of demonstrating defect ordering. The relaxation rate of the polarization is set by the rotational diffusion $D_R$ of the $+1/2$ defect and it receives nonlinear passive corrections proportional to $\rho$ and $|\b{v}_n|^2$ with $\beta=5\nu^2\kappa/8$, as described by the last two terms in square brackets. Note that, although the defect charge density $\rho$ can be locally negative, this doesn't lead to an instability in the polarization equation at equilibrium, because $\langle\rho\rangle=0$ in a charge neutral system and deviations from this mean value relax on a finite timescale. The convective-like term $\hat{\b{z}}\cdot(\b{v}_n\times\bm\del)\b{p}$ accounts for the change in the polarization due to the passive motion of $+1/2$ defects and the penultimate term on the right hand side of Eq.~\ref{eq:peq} is a passive elastic torque that rotates the defect polarization in response to elastic distortions (see also Appendix~\ref{app:poldyn}). Activity enters in two places: in the pressure like term $\sim (v_1/2)\bm\del n_+$ and as an active torque $\sim v_2n_+\,\vec{\e}\cdot\b{v}_n$ that builds up local polar order due to transverse director deformations. The active coefficients $v_1$ and $v_2$ are defined as
\begin{gather}
	v_1=v\dfrac{2T}{K}f(\zeta)\ ,\quad v_2=v\dfrac{3\nu T}{4K}f(\zeta)\ ,\label{eq:v1v2}\\
	\zeta=\dfrac{v^2\gamma T}{D_RK^2}\ ,\label{eq:zeta}
\end{gather}
where $\zeta$ is a nondimensional activity. The positive dimensionless function $f(\zeta)=1+4\zeta+\mathcal{O}(\zeta^2)$ captures the nonlinear dependence of the torque on activity and can be computed from the moment hierarchy (see Appendix~\ref{app:calc}). Below we shall analyze the steady states of these equations and their stabilty.

\section{Homogeneous states and phase transitions}
\label{sec:states}
We consider a state with a homogeneous density of unbound disclinations. By charge neutrality in the plane and the topological constraint (Eq.~\ref{eq:topo}), we have $\rho=0$ and hence $n_+=n_-=n$.
Setting all the gradient terms to zero, we obtain
\begin{align}
	\partial_tn&=W_c-W_a\ ,\\
	\partial_t\b{v}_n&=\pi v\,\vec{\e}\cdot\b{p}-2\pi\nu\kappa\,n\,\b{v}_n\ ,\label{eq:vshomo}\\
	\partial_t\b{p}&=-\left[D_R+\beta|\b{v}_n|^2\right]\b{p}-v_2n\,\vec{\e}\cdot\b{v}_n\ .\label{eq:phomo}
\end{align}
At steady state, we set $n=n_0$ such that $W_c=W_a$, $\b{v}_n=\b{v}_n^0$ and $\b{p}=\b{p}_0$. From Eq.~\ref{eq:vshomo} we have $\b{v}_n^{0}=(v/2\nu\kappa n_0)\vec{\e}\cdot\b{p}_0$, which corresponds to a vanishing charge current ($\b{j}_{\rho}=\b{0}$). Eliminating $\b{v}_n^0$ from Eq.~\ref{eq:phomo} and after some algebraic manipulations we get
\begin{equation}
	D_R\left[a_2-a_4|\b{p}_0|^2\right]\b{p}_0=\b{0}\ ,\label{eq:porder}
\end{equation}
where 
\begin{equation}
\label{eq:a2a4}
	a_2=\dfrac{3}{8}\zeta f(\zeta)-1\ ,\quad a_4=\zeta\dfrac{5K}{32Tn_0^2}\ .
\end{equation}
Note that $a_2$ can change sign at high activity ($\zeta\sim v^2$) and $a_4>0$. For small activity $a_2<0$ and the only solution is $\b{p}_0=\b{0}$ and $\b{v}_n^0=\b{0}$. At large activity $a_2>0$ and we obtain a solution with $|\b{p}_0|\neq 0$ and therefore $|\b{v}_n^0|\neq 0$, corresponding to a uniformly polarized defect ordered state. The change in sign of $a_2$  occurs at a critical value of activity set by
\begin{equation}
	\zeta_c\,f(\zeta_c)=\dfrac{8}{3}\implies\zeta_c\approx0.701\ ,
	\label{eq:zetac}
\end{equation}
where we have used the leading order expansion of $f(\zeta)=1+4\zeta+\mathcal{O}(\zeta^2)$. Including higher order terms in the expansion of $f(\zeta)$ will change the numerical value of $\zeta_c$, which will, however, remain finite.
 Altogether we have three distinct homogeneous steady states:
\begin{itemize}
	\item Homogeneous nematic order with no unbound disclinations ($n_0=0$, $\b{v}_n^0=\b{p}_0=\b{0}$).
	\item Isotropic defect chaos ($n_0>0$, $\b{v}_n^0=\b{p}_0=\b{0}$).
	\item Defect polar ordered state ($n_0>0$, $\b{v}_n^0,\,\b{p}_0\neq\b{0}$).
\end{itemize}
The three phases along with the intervening phase boundaries discussed below are shown in the phase diagram in Fig.~\ref{fig:phasedgm}, as a function of $v$, $T$ and $D_R$, keeping $\nu$ fixed. Changing $\nu$ does not change the global topology of the phase diagram, but affects the phase boundaries only in a quantitative way. 

\subsection{Defect unbinding transition}
The nematic order-disorder transition is mediated by an activity driven unbinding of defect pairs \cite{shankar2018defect}. The resulting defect-ridden state is isotropic and disordered ($\b{p}_0=\b{0}$) with a finite nematic correlation length $\xi\sim1/\sqrt{n_0}$.
Given the disordered motion of the $+1/2$ disclinations, we identify this state with the spatio-temporally chaotic dynamics of ``active turbulence''. The activity threshold for defect unbinding was obtained in Ref.~\cite{shankar2018defect} to be
\begin{equation}
	|v_{c_1}|=\sqrt{\dfrac{2\nu\kappa\,D_R\left(1-\tilde{T}\right)}{\pi\left[1+3\nu\,\tilde{T}/32\right]}}\ ,\label{eq:vc1}
\end{equation}
where $\tilde{T}=T/T_c^{\mathrm{eq}}$ is a normalized effective temperature with $T_c^{\mathrm{eq}}=\pi K/8$ the equilibrium Kosterlitz-Thouless transition temperature \cite{kosterlitz1973ordering,stein1978kosterlitz}. This transition line is marked in blue in Fig.~\ref{fig:phasedgm}.

The location of the defect unbinding transition can be understood by a simple argument first given in Ref.~\cite{shankar2018defect} and repeated here for completeness. In a na{\"i}ve one-dimensional picture, where the two defects of a neutral pair ($\pm1/2$) unbind by moving away from each other along a straight line, the self propulsion of the $+1/2$ defect can always overcome the passive Coulomb attraction, resulting in defect unbinding at any activity. On the other hand, rotational diffusion ($D_R$) can spoil this process by endowing the $+1/2$ defect with a finite persistence length $\ell_p=|v|/D_R$. When $\ell_p<r_c$, with $r_c\sim\mu K/|v|$ being the pair separation where the propulsive force $|v|/\mu$ and the attractive Coulomb forces $K/r$ balance, rotational noise disrupts the straight path of the $+1/2$ defect before it can overcome the energy barrier required for unbinding, allowing defect pairs to remain bound. The condition for unbinding due to the activity can then be estimated as $\ell_p\sim r_c$, which coincides with Eq.~\ref{eq:vc1} for small $T$.

\subsection{Defect ordering transition}
\label{subsec:defectorder}
For $\zeta>\zeta_c$ as given by Eq.~\ref{eq:zetac} the isotropic gas of defects spontaneously breaks rotational symmetry by ordering into a polar, collectively moving liquid. Expressing the transition point  in terms of the original model parameters, we find that the defect ordering transition occurs at
\begin{equation}
	|v_{c_2}|=\sqrt{\zeta_c\dfrac{D_RK^2}{\gamma T}}\simeq0.84\sqrt{\dfrac{D_RK^2}{\gamma T}}\ ,
\end{equation}
which is shown as a red line in Fig.~\ref{fig:phasedgm}.

A simple yet physical way to understand this threshold for defect ordering is as follows. Disregarding the numerical constant $\zeta_c$, the condition for defect ordering ($|v|>|v_{c_2}|$) can be written as
\begin{equation}
	\left(\dfrac{T}{K}\right)\dfrac{\gamma\ell_p^2}{K}\gtrsim D_R^{-1}\ .
\end{equation}
The factor $T/K$ accounts for the relative strength of fluctuations $\sim T$ to the defect core energy $\sim K$ \cite{chaikin2000principles}, and it corresponds to the cost to nucleate a defect pair. Once created, the $+1/2$ disclination self-propels itself away from the $-1/2$ disclination, quasi-deterministically on time scales shorter than the rotational diffusion time $D_R^{-1}$. In doing so, it distorts the underlying nematic in its wake, on the scale of a defect persistence length $\ell_p$. As $\gamma\ell_p^2/K$ is the time it takes the underlying nematic to relax distortions on a length scale of $\ell_p$, the threshold for defect ordering is a simple balance of time scales.

When the nematic rapidly heals the distortion created by the swarming $+1/2$ disclination, faster than the defect reorientation time ($\gamma\ell_p^2/K\ll D_R^{-1}$), we obtain an isotropic disordered state of defect chaos. In the opposite limit, the underlying nematic responds too slowly and is unable to relax the distortion left by the active defects. This leads to the formation of a long-lived kink wall (also called a $\pi$- or N{\'e}el-wall) that terminates at the defect pair. This locally frozen-in distortion feeds back into the defect motion, leading to a build up of polar order through a combination of many-body screening and active torques. Eventually, for strong enough activity, the underlying nematic cannot catch up with the persistent dynamics of the defects, which then condense into a spontaneously flowing defect polarized liquid, i.e., a $+1/2$ defect flock. The active self-aligning torques derived in Ref.~\cite{shankar2018defect} are a crucial ingredient to this mechanism of defect ordering. Reminding ourselves of the physics of the active torque, we note that it arises from the active backflow induced defect motion advecting the defect polarization itself. This results in an effective torque which tries to align the defect orientation to its velocity, with a magnitude controlled directly by activity.
While similar torques with short ranged interaction forces have been used to model flocking in both cells \cite{szabo2006phase,henkes2011active} and vibrated polar grains \cite{weber2013long,lam2015self}, it is only in the presence of many-body screening that the active torques can cause collective motion of defects in active nematics.

Close to the defect ordering transition, for $\zeta>\zeta_c$ we have
\begin{equation}
	|\b{p}_0|\simeq n_0\sqrt{\dfrac{T}{K}}\left(\dfrac{\zeta-\zeta_c}{\zeta_c}\right)^{1/2}\ ,
\end{equation}
with the usual mean-field exponent, although fluctuations are expected to decrease it. As the phase gradient $\b{v}_n=\bm\del\theta$ is slaved to the polar order, we also have
\begin{equation}
	\b{v}_n^0=\dfrac{v}{2\nu\kappa n_0}\vec{\e}\cdot\b{p}_0\propto\left(\dfrac{\zeta-\zeta_c}{\zeta_c}\right)^{1/2}\ ,
\end{equation}
with $\b{p}_0\cdot\b{v}_n^0=0$.
The appearance of polar defect order also spontaneously breaks \emph{translational} symmetry of the underlying nematic in the direction orthogonal to that of defect order, with $\theta(\b{r})\simeq \theta_0+\b{v}_n^0\cdot\b{r}$. While the defect liquid itself has no translational order, the underlying nematic undergoes a concomitant modulational instability and develops a smectic array of splay-bend kink walls, similar to that seen in numerical simulations \cite{decamp2015orientational,putzig2016instabilities,srivastava2016negative,patelli2019understanding}. A cartoon of the defect and kink wall structure in the defect ordered state is sketched in Fig.~\ref{fig:flock}.

\begin{figure}[]
	\centering
	\includegraphics[width=0.38\textwidth]{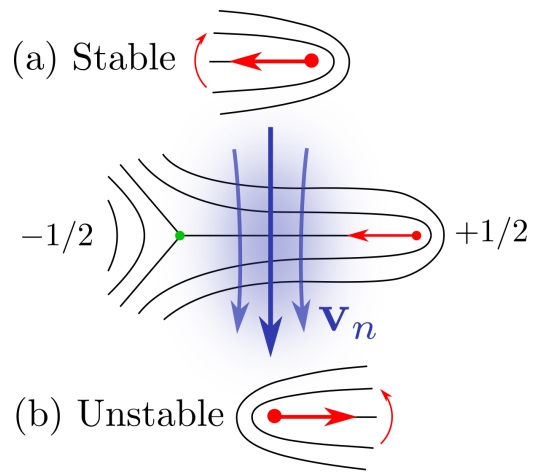}
	\caption{The orientational stability of a test $+1/2$ disclination placed in the vicinity of a neutral $\pm1/2$ defect pair, shown here for extensile activity ($v<0$). The $\pm1/2$ defect pair creates a background phase gradient $\b{v}_n$ (shown as blue arrows) as a result of the nematic distortion. The test disclination experiences an active torque due to the presence of a background phase gradient $\b{v}_n$ that picks out a stable orientation as shown in (a). The torque-stabilized orientation is one in which the test defect is aligned parallel to the defect pair. All other orientations of the test defect, including being antiparallel to the defect pair as in (b), are unstable due to the active torque.}
	\label{fig:stability}
\end{figure}

Finally, in light of the apolar nature of an active nematic, the appearance of a directed polar current might seem surprising. The polarization density is an emergent property of nonlinear topological excitations in the system which then permits spontaneous flow due to the absence of detailed balance. On the other hand, experiments in microtubule suspensions have reported  \emph{apolar} (i.e., nematic) ordering of $+1/2$ disclinations~\cite{decamp2015orientational} that may seem a more natural possibility. Nematic defect order is not possible in our model due to the nature of defect interactions, particularly the active torques. To see this, consider a test $+1/2$ disclination placed in the vicinity of a neutral defect pair. The $\pm1/2$ defect pair instantaneously creates a background phase gradient $\b{v}_n$ orthogonal to the line joining the two defect cores as shown in Fig.~\ref{fig:stability}. When a test $+1/2$ disclination is placed in this background distortion, one immediately sees that the active torque preferentially stabilizes its orientation to be aligned \emph{with} the $+1/2$ defect in the neutral pair. This is illustrated in Fig.~\ref{fig:stability} for the case of extensile activity ($v<0$) and is discussed in more detail in Appendix~\ref{app:efield}. The contractile case works analogously. Passive elastic torques $\sim\kappa\bm\del\cdot\b{v}_n$ cannot change this as they vanish for spatially homogeneous phase gradients, unlike the active torques. Hence active torques always induce an effective \emph{polar} alignment between $+1/2$ defects, thereby structurally preventing the appearance of nematic defect order. This should be contrasted with other examples of defect order in equilibrium - the Abrikosov vortex lattice in type II superconducting films \cite{abrikosov1957magnetic} and the twist-grain boundary phase in smectics \cite{renn1988abrikosov}, both of which result from the local breakdown of Meissner like effects, leading to the penetration of either the magnetic field or twist, respectively.


\section{Fluctuations and linear stability}
\label{sec:stability}
We shall now examine the linear stability of the homogeneous steady states to small spatial fluctuations. In both the isotropic and defect ordered states, fluctuations in the average defect number density ($n$) relax on a short time scale set by the balance of defect pair creation and annihilation and are henceforth neglected. In the presence of unbound defects, the nematic order parameter has a finite correlation length $\xi$ which sets the mean separation between defects. Fixing $n=n_0$ we define
\begin{equation}
	\xi=\dfrac{1}{\sqrt{2\pi\nu n_0}}\ .
\end{equation}
The numerical factor of $2\pi\nu$ is introduced to simplify the notation below.
\subsection{Isotropic defect chaos}
\label{subsec:chaos}
Linearizing for small fluctuations about the isotropic steady state, we have $\rho=0+\delta\rho$, $\b{v}_n=\b{0}+\delta\b{v}_n$, $\b{p}=\b{0}+\delta\b{p}$.
Upon Fourier transforming in space ($\Phi_{\b{q},\omega}=\int\dd^2r\int\dd t\ e^{i\omega t-i\b{q}\cdot\b{r}}\Phi(\b{r},t)$), we define the longitudinal and transverse components of $\delta\b{p}$ as $\delta p_L=\hat{\b{q}}\cdot\delta\b{p}_{\b{q}}$ and $\delta p_T=\hat{\b{z}}\cdot\left(\hat{\b{q}}\times\delta\b{p}_{\b{q}}\right)$ respectively. A similar decomposition is performed for $\delta\b{v}_n$ as well. The resulting linearized equations are
\begin{align}
	\partial_t\delta p_L&=-D_R\left(1+q^2\ell^2\right)\delta p_L-v_2n_0\left(1-\dfrac{4}{3}q^2\xi^2\right)\delta\mathrm{v}_n^T\ ,\\
	\partial_t\delta\mathrm{v}_n^T&=-\dfrac{1}{\tau}\left(1+\dfrac{D_0}{\kappa}q^2\xi^2\right)\delta\mathrm{v}_n^T-\pi v\,\delta p_L\ ,\\
	\partial_t\delta p_T&=-D_R\left(1+q^2\ell^2\right)\delta p_T+v_2n_0\,\delta\mathrm{v}_n^L\ ,\\
	\partial_t\delta\mathrm{v}_n^L&=-\dfrac{1}{\tau}\left(1+q^2\xi^2\right)\delta\mathrm{v}_n^L+\pi v\,\delta p_T\ .
\end{align}
Note that the four fields are only coupled in pairs; $(\delta p_L,\,v_n^T)$ and $(\delta p_T,\,v_n^L)$. Here we have used the topological constraint $\hat{\b{z}}\cdot\left(\bm\del\times\delta\b{v}_n\right)=2\pi\delta\rho$ and introduced $\tau=(2\pi\nu\kappa n_0)^{-1}=\xi^2/\kappa$ as the finite relaxation time of the phase gradient. The thermal diffusion length scale $\ell^2=D_0/D_R$ is typically microscopic and expected to be of the order of the defect core size ($\ell\sim a$). Although the charge density $\rho$ is locally conserved (Eq.~\ref{eq:rhoeq}), its fluctuations are slaved entirely to $\delta\mathrm{v}_n^T$ which decays on a finite time scale $\tau$, due to complete screening of the long-ranged Coulomb (passive) interaction between the defects.

Both pairs of coupled modes have the same dispersion relation at $\mathcal{O}(q^0)$, given by
\begin{equation}
	i\omega_{\pm}=\dfrac{1}{2}\left[\dfrac{1}{\tau}+D_R\pm\sqrt{\left(D_R-\dfrac{1}{\tau}\right)^2+\dfrac{3D_R}{2\tau}\zeta f(\zeta)}\right]\ ,\label{eq:omega}
\end{equation}
differing from $\mathcal{O}(q^2)$ terms onwards. While $i\omega_+>0$ always, $i\omega_-$ can go negative for large enough activity ($\zeta$) triggering an instability of the isotropic state. This instability coincides with  the defect ordering transition at  $\zeta=\zeta_c$ (obtained in Eq.~\ref{eq:zetac}). One can check that $\b{q}$-dependent terms do not change this and no further instabilities arise.

As we approach the defect unbinding transition from above, within the defect chaos state, the nematic correlation length diverges ($\xi\to\infty$) and the density of free defects vanishes ($n_0\to 0$). Polarization fluctuations relax with a finite rate due to rotational diffusion, even at the unbinding transition ($i\omega_+\simeq D_R$), while the phase gradient exhibits critical slowing down with $i\omega_-\propto\xi^{-2}\to 0$. Of course the region of hydrodynamic validity $q\xi\ll 1$ shrinks rapidly as we approach the defect unbinding transition.

Reinstating the lowest order additive noise as computed from coarse-graining ($2\pi\sqrt{D_0n_0}\,\vec{\Lambda}_1$ in Eq.~\ref{eq:vseq} and $\sqrt{2TD_Rn_0f(\zeta)/K}\,\vec{\Lambda}_2$ in Eq.~\ref{eq:peq}, with both $\vec{\Lambda}_1$ and $\vec{\Lambda}_2$ uncorrelated unit white Gaussian noise), we compute the fluctuation spectra in the isotropic defect gas. Using $q=|\b{q}|$, the equal time correlator of the polarization at steady state is given by,
\begin{equation}
	\langle|\delta\b{p}_{\b{q}}(t)|^2\rangle\approx n_0\dfrac{2T}{K}f(\zeta)+\dfrac{n_0D_0\,(v_2/\nu\kappa)^2}{2D_R\left(1+q^2\xi^2\right)\left[1+\tau D_R+q^2\xi^2\right]}\ .\label{eq:pcorr}
\end{equation}
As we are far from the defect ordering threshold ($\zeta\ll\zeta_c$), we have only retained the most dominant terms involving $\delta p_T$ and neglected $\ell\ll\xi$ for simplicity. In this limit, $\delta p_L$ has correlations only on scales much smaller than $\xi$ and mainly contributes an additive constant in Eq.~\ref{eq:pcorr}. Note the dependence on $n_0$ that arises from the variance of the noise in the polarization dynamics. The polarization inherits its dominant spatial correlations from $\b{v}_n$, which survives on the scale of $\xi$. Note that this results in a breakdown of dynamic scaling close to the defect unbinding transition. The charge density fluctuations for $\b{q}\to\b{0}$ are similarly given by
\begin{equation}
	\langle|\delta\rho_{\b{q}}(t)|^2\rangle\approx\dfrac{q^2}{4\pi\nu\kappa}\left[D_0+\left(\dfrac{\tau D_R}{1+\tau D_R}\right)\kappa\,\zeta f(\zeta)\right]\ ,
\end{equation}
upon assuming $\zeta\ll\zeta_c$, well below the defect ordering threshold.
Charge density fluctuations vanish on large scales as a consequence of the unbound defect gas behaving as a screened conducting plasma albeit with an activity enhanced effective ``dielectric constant'' via a sum rule \cite{martin1988sum}. While such a dielectric constant relates to an effective elastic constant or stiffness in 2D equilibrium superfluids or XY magnets, the equivalent correspondence in the active case requires care and will be addressed elsewhere \cite{shankar2020}.

A common feature of defect chaos is the characteristic decay of the flow velocity correlator and the kinetic energy spectrum. We assume that the defects serve as faithful tracers of the average flow field and write
\begin{equation}
	\b{u}=\dfrac{\b{j}_n}{n}\approx\dfrac{v}{2n_0}\delta\b{p}\ ,\label{eq:u}
\end{equation}
where the last approximate equality is obtained by linearizing $\b{j}_n$ for small $\delta\b{p}$ and $\delta\b{v}_n$ and neglecting higher order gradient terms. This approximation of course breaks down close to the defect unbinding transition when $n_0\to 0$.

Two distinguishing properties are generally associated with ``active turbulence''. The first is that the typical length scale of the flow depends on the magnitude of activity $|\alpha|$ through the mean defect spacing $\xi$, with $\xi\sim |\alpha|^{-1/2}$ \cite{sanchez2012spontaneous,thampi2013velocity,giomi2015geometry,hemingway2016correlation,lemma2018statistical}. The second is the scaling of the kinetic energy spectrum $E(\b{q})=\langle|\b{u}_{\b{q}}|^2\rangle/2$, with $E(q\to0)\propto|\alpha|$. In our work $|v|\sim|\alpha|$ (see Appendix~\ref{app:calc}) and $\xi\sim n_0^{-1/2}$ away from the defect unbinding threshold. Deep in the regime of defect chaos, the mean density of defects is known to scale with the magnitude of activity \cite{giomi2015geometry,hemingway2016correlation}, hence $\xi\sim |v|^{-1/2}\sim|\alpha|^{-1/2}$. To derive this scaling from semi-microscopic considerations requires a full theory of melting, far beyond the scope of this paper, and will be presented elsewhere \cite{shankar2020}. As can be seen directly from Eq.~\ref{eq:u} and Eq.~\ref{eq:pcorr}, in our model the characteristic length scale of both flow and defect polarization indeed scales as $\xi$ far from the defect unbinding threshold. For the second point, we can obtain the scaling of the kinetic energy using Eq.~\ref{eq:u} and \ref{eq:pcorr}, with the result
$E(q\to0)\propto|v|(1+\mathcal{O}(|v|))$. The leading $|v|$ scaling of the average kinetic energy crucially relies on the fact that the polarization fluctuations $\propto n_0$ (Eq.~\ref{eq:pcorr}), a consequence of the central limit theorem. Both these scalings no longer hold near  the defect unbinding transition. It is worth emphasizing that our calculation based on a systematic derivation of the defect dynamics is consistent with the phenomenological mean-field picture of Ref.~\cite{giomi2015geometry}. The  two approaches are complementary, with the mean-field construction working best on short scales where correlations are neglected, while our approach works best on large scales where hydrodynamic treatments are applicable.

\subsection{Defect polar order}
\label{subsec:polar}
Deep in the defect ordered state, both $\b{v}_n$ and $\rho$ are fast modes and rapidly relax to their steady-state values ($\b{v}_n^0$ and $0$ respectively) on the now short time scale $\tau$. It is instructive (although not essential) to slave both these fields to the polarization. In doing so, we set $\b{v}_n=\pi v\tau\,\vec{\e}\cdot\b{p}+\mathcal{O}(\bm\del)$ and $\rho=-(v\tau/2)\bm\del\cdot\b{p}+\mathcal{O}(\del^2)$ to obtain an effective Toner-Tu like equation for the defect polarization,
\begin{align}
	\partial_t\b{p}+&\lambda_1\b{p}\cdot\bm\del\b{p}+\lambda_2\b{p}\bm\del\cdot\b{p}+\lambda_3\bm\del|\b{p}|^2\nonumber\\
	=&D_R\left[a_2-a_4\,\b{p}^2\right]\b{p}+D_0\del^2\b{p}+D_1\bm\del\bm\del\cdot\b{p}\ .\label{eq:tonertu}
\end{align}
The resulting hydrodynamic coefficients are
\begin{gather}
	\lambda_1=-\dfrac{v}{2\nu n_0}(2+\nu)\ ,\ \lambda_2=-\dfrac{v}{2n_0}\left(1-\dfrac{vv_2\tau}{4\nu\kappa}\right)\label{eq:coeff1}\\
	\lambda_3=\dfrac{v}{2\nu n_0}\ , \ D_1=\dfrac{vv_1\tau}{4}\ .\label{eq:coeff2}
\end{gather}
with $a_2,a_4$ given in Eq.~\ref{eq:a2a4}. Note that, to leading order as $n_0\propto|v|$ ($|v|\sim|\alpha|$), we find that $\lambda_{1,2,3}\sim\mathrm{sgn}(v)$ is nonanalytic in activity, and the effective splay elastic constant $D_1\sim|v|/K$ is controlled by the well-known active length scale \cite{doostmohammadi2018active}.
As noted earlier in Sec.~\ref{subsec:defectorder}, for $\zeta>\zeta_c\simeq0.701$, $a_2>0$ and the isotropic defect gas spontaneously orders into a polarized liquid. The finite relaxation time for the defect charge density $\rho$ implies that the defect ordered liquid behaves as a Malthusian flock \cite{toner2012birth}. From Eq.~\ref{eq:tonertu}, we also see that the important convective nonlinearity $\lambda_1\b{p}\cdot\bm\del\b{p}$ naturally appears in our framework and it allows the existence of \emph{long-ranged} polar order of the $+1/2$ disclinations. The spontaneous breaking of rotational symmetry is accompanied by a simultaneous breaking of translational symmetry in the underlying nematic, characterized by the appearance of periodic kink walls with a nonzero average phase gradient $\langle\b{v}_n\rangle=\b{v}_n^0$. In Fig.~\ref{fig:flock}, we see that the $+1/2$ defects preferentially move along the kink walls, suggesting a structure akin to the ``active smectic-P'' state recently reported in Ref.~\cite{romanczuk2016emergent}. This superficial similarity goes no further as neither the defect number nor charge density reflect the necessary periodic modulation. This is an important distinction, since the periodic arrangement of kink walls and associated \emph{long-ranged} smectic order (unlike all other 2D active smectics \cite{adhyapak2013live,chen2013universality,romanczuk2016emergent}) is not an independent broken symmetry, but rather the result of the LRO of the polarization. The latter in turn arises as in all Toner-Tu models from the convective nonlinearity in Eq.~(\ref{eq:tonertu}).

Linearizing for small fluctuations about the defect ordered steady state, we only have $\b{p}=\b{p}_0 +\delta\b{p}$ as fluctuations in both $\b{v}_n$ and $\rho$ have already been enslaved. Decomposing $\delta\b{p}$ along ($\delta p_{||}$) and transverse ($\delta p_{\perp}$) to the polar order, we obtain two coupled modes whose long-wavelength dispersion relation is of the form
\begin{equation}
	i\omega_{||,\perp}(\b{q})=ic_{||,\perp}\,q_{||}+\Gamma_{||,\perp}(\b{q})\ ,
\end{equation}
where $q_{||}=\b{q}\cdot\hat{\b{p}}_0$. Both amplitude ($\delta p_{||}$) and orientational ($\delta p_{\perp}$) fluctuations propagate along the direction of polar order, though with different speeds,
\begin{equation}
	c_{||}=(\lambda_1+\lambda_2+2\lambda_3)p_0\ ,\ c_{\perp}=\lambda_1p_0\ ,
\end{equation}
having used the magnitude of polarization $|\b{p}_0|=p_0$. Depending on the activity ($v$), these drift speeds can be of either sign (see Eqs.~\ref{eq:coeff1},~\ref{eq:coeff2}). The relaxation rates for the two modes upto $\mathcal{O}(q^2)$ are given as
\begin{gather}
	\Gamma_{||}(\b{q})=2a_4D_Rp_0^2+\left(D_0+D_1\cos^2\phi-\dfrac{\lambda_2\lambda_3}{a_4D_R}\sin^2\phi\right)q^2\ ,\label{eq:gamma||}\\
	\Gamma_{\perp}(\b{q})=\left[D_0+\left(D_1+\dfrac{\lambda_2\lambda_3}{a_4D_R}\right)\sin^2\phi\right]q^2\ ,
\end{gather}
where $\phi$ is the angle made by the wave-vector $\b{q}$ with the direction of polar order ($\hat{\b{q}}\cdot\hat{\b{p}}_0=\cos\phi$). As expected, in the defect ordered state, the amplitude mode decays with a finite relaxation rate even as $\b{q}\to\b{0}$, while orientational fluctuations remain soft. Being a broken symmetry variable, this is the only true hydrodynamic mode here. From our derivation of transport coefficients (Eqs.~\ref{eq:coeff1},~\ref{eq:coeff2}), we note that $\lambda_2\lambda_3<0$ for small $v$, changing sign for larger activity. This is true for both extensile and contractile systems. Hence, while the amplitude mode is always stable ($\Gamma_{||}(\b{q})>0$) \footnote{Though, from Eq.~\ref{eq:gamma||}, it might seem that $\delta p_{||}$ can go unstable at high activity when $\lambda_2\lambda_3>0$, this is an artifact of the low $\b{q}$ expansion. In fact, for $\phi=\pi/2$, the complete dispersion relation for the amplitude mode is $i\omega_{||}=[\tilde{a}_4+(2D_0+D_1)q^2+\sqrt{(\tilde{a}_4-D_1q^2)^2-4q^2\tilde{a}_4\Lambda}]/2$, where $\tilde{a}_4=2a_4D_Rp_0^2$ and $\Lambda=\lambda_2\lambda_3/a_4D_R$. For either sign of $\Lambda$, $\mathrm{Re}(i\omega_{||})>0$ for all $q$, confirming that the only possible instability is the orientational one obtained in Eq.~\ref{eq:splay}} the orientational mode $\delta p_{\perp}$ can develop a long-wavelength splay instability (at $\phi=\pi/2$) if $\Gamma_{\perp}(\b{q})<0$. Rewriting this condition in terms of our original control parameters, this corresponds to
\begin{equation}
	D_0+\dfrac{\zeta f(\zeta)D_R}{4\pi\nu\,n_0}\left(1+\dfrac{3}{5\nu}\right)-\dfrac{8\kappa}{5\nu}<0\ ,\label{eq:splay}
\end{equation}
for a splay instability to occur. The detailed dependence of $n_0$ on system parameters is sensitive to microscopic details, so this instability threshold is model dependent. As $\zeta\propto v^2$ and $n_0\propto |v|$ along with $f(\zeta)$ monotonic and positive, for large enough activity, we expect the above condition to \emph{not} be satisfied and thereby allow a stable polar ordered phase of defects. In this paper we only consider the stable situation and shall not discuss the splay instability further.

Putting back noise as before, we can compute the correlation function of the polarization fluctuations. The convective term is a relevant nonlinearity in 2D that dramatically modifies the scaling of the autocorrelation, and is well known for being responsible in stabilizing long-ranged polar order. Using the exact exponents calculated in Ref.~\cite{toner2012birth}, we have at steady-state
\begin{equation}
	\langle\delta p_{\perp}(\b{r},t)\delta p_{\perp}(\b{0},0)\rangle=r_{\perp}^{-2/5}G\left(\dfrac{r_{||}-c_{\perp}t}{r_{\perp}^{3/5}},\dfrac{t}{r_{\perp}^{6/5}}\right)\ .
\end{equation}
As before we have used $r_{||}=\hat{\b{p}}_0\cdot\b{r}$ and $r_{\perp}=\hat{\b{z}}\cdot\left(\hat{\b{p}}_0\times\b{r}\right)$, and $G(x,y)$ is a scaling function that depends on model parameters in a complicated way.
Both defect charge and number density exhibit normal Poissonian fluctuations, as expected of a Malthusian flock. The defect ordered liquid continually turns over due to spontaneous pair creation and annihilation of defects, with polar order persisting for infinitely longer than the finite lifetime of individual disclinations. The conspicuous absence of both giant number fluctuations \cite{ramaswamy2003active,narayan2007long,marchetti2013hydrodynamics} and motility-induced phase separation phenomenology \cite{tailleur2008statistical,fily2012athermal,cates2015motility} for defects is at first glance surprising. The local conservation of topological charge, along with the self-propulsion of $+1/2$ disclinations might lead one to na{\"i}vely expect large fluctuations in the charge density upon defect ordering. In addition, the repulsive interaction between motile $+1/2$ disclinations might also raise the question of the possibility of phase separation of a defect liquid. None of these scenarios are realized. The fundamental reason is the nature of the screened Coulomb interaction between defects mediated by Frank elasticity, which renders the charge density a non-hydrodynamic field with a finite relaxation time.

This concludes the analysis of the phase diagram predicted by our model of defect hydrodynamics. Untill now, all the phases considered had no charge separation and the average defect charge current always vanished ($\langle\b{j}_{\rho}\rangle=\b{0}$). In 2D passive systems such as superfluid films, a charge current can be generated by the application of an external electric field. While applying a similar external field in active nematics might be tricky, we will show below that spatially varying activity can locally act as an ``electric field'' causing a local sorting of defects based on their topological charge.

\section{Inhomogeneous activity}
\label{sec:inhomo}
\begin{figure*}[]
	\centering
	\includegraphics[width=0.8\textwidth]{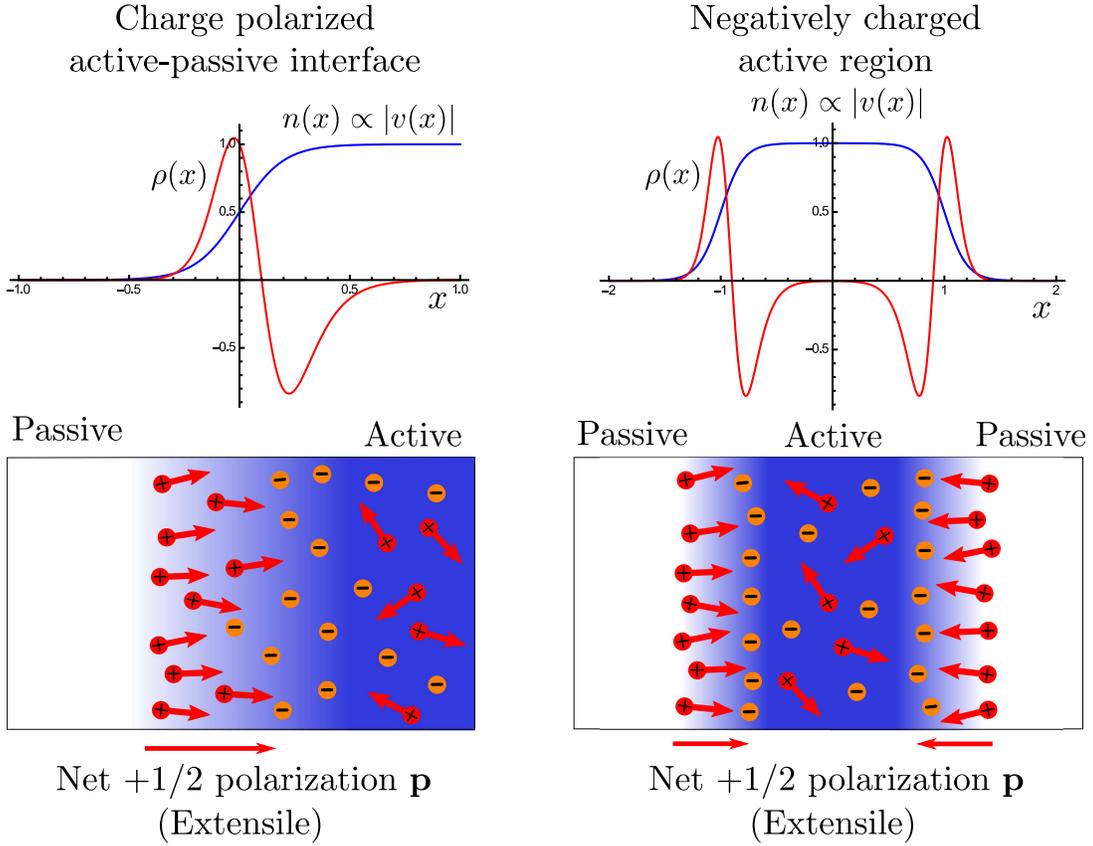}
	\caption{The defect charge and number density along with the polarization in a 2D nematic with active-passive interfaces. The activity profile is simply chosen to be $v(x)=(v/2)[1+\tanh(2x/w)]$ for a passive ($x<0$) to active ($x>0$) interface, $v$ being the maximum activity and $w$ is the width of the interface. As can be seen, the interface develops a steady-state charge polarization, with excess $+1/2$ disclinations near the passive side and a compensating number of $-1/2$ disclinations near the active side. Within the bulk active region with isotropic defect chaos, the polarization vanishes, but at the interface, the $+1/2$ disclinations locally order to orientationally polarize the interface as well. The direction of interfacial polarization is shown in the cartoon for the extensile case ($v<0$), with the defects just marked by the sign of their topological charges.}
	\label{fig:interface}
\end{figure*}
We now demonstrate that our hydrodynamic model provides a useful framework for describing situations with spatially inhomogeneous profiles of activity. Activity gradients provide additional nonequilibrium driving forces that can be used to generate and control spatial patterns at will. Recent advances in engineering optical control of biomolecular or catalytic activity have emerged as a venerable platform for dynamically creating large scale reconfigurable patterns and structures in diverse systems ranging from gels of biofilament-motor complexes \cite{schuppler2016boundaries,ross2018controlling} to suspensions of self-propelled colloids \cite{palacci2013living} and bacteria \cite{arlt2018painting,frangipane2018dynamic}. Such techniques, when used in active nematics, can control the chaotic dynamics and provide novel ways to precisely sculpt flow and tune material properties through the spatial organization and positioning of topological defects. This is an important pre-requisite for developing programmable active metamaterials whose local organization dictates its global response and transport.

To include spatially inhomogeneous activity in our defect hydrodynamics, we make use of the fact that activity enters through the defect self-propulsion $v$ and simply replace $v\to v(\b{r})$ (including it under gradients when appropriate). While modulating the microscopic active processes will in general also affect the elasticity and viscosity of the nematic, we shall not consider such modifications for simplicity. There are two distinct phenomena that result when activity is spatially varying in an active nematic. First, activity controls the motility of $+1/2$ disclinations, which behave as active particles that are known to aggregate where they move slowly \cite{schnitzer1993theory,tailleur2008statistical,cates2015motility}. This well understood phenomenon survives even in the presence of interactions and has been used to design self-assembled rectification devices \cite{stenhammar2016light} and particle traps \cite{magiera2015trapping,sharma2017brownian,grauer2018spontaneous}. The second phenomenon is a distinct property of active nematics, which is that higher activity generates more defect pairs. Hence, low activity leads to low defect motility and a consequent accumulation of $+1/2$ defects, but it also decreases the total defect density. As the two competing effects don't act symmetrically on both charge defects, we have the possibility of sorting defects by charge in the presence of an activity gradient.

Working for simplicity in a one-dimensional (1D) setting, we consider activity to only vary in the $x$-direction ($v(x)$). We will take the maximum value of activity to correspond to states deep in the regime of defect chaos and to never exceed the defect ordering threshold. As in Sec.~\ref{subsec:chaos}, we assume that the average defect density relaxes to its steady state value on a short time scale \footnote{This neglects changes in $n$ due to a defect flux $\bm\del\cdot\b{j}_n$, but it can be accounted for easily.}, and all other fields are small with weak gradients, permitting a linearized analysis. This is done simply for analytical progress - alternatively the equations could be solved numerically. The dependence of $W_c$ and $W_a$ on $v(x)$ causes the average defect number density to also have a spatial profile $n=n_0(x)\propto|v(x)|$. Due to the 1D setup, only $\b{p}=p(x)\hat{\b{x}}$ and $\b{v}_n=\mathrm{v}_n(x)\hat{\b{y}}$ are nonvanishing, and at steady state, once again $\b{j}_{\rho}=\b{0}$. The linearized steady state equations to leading order in gradients are
\begin{gather}
	\dfrac{v(x)}{2}p(x)+\nu\kappa\,n_0(x)\mathrm{v}_n(x)=0\ ,\\
	\dfrac{1}{2}\partial_x\left[v_1(x)n_0(x)\right]=-D_Rp(x)-v_2(x)n_0(x)\mathrm{v}_n(x)\ .
\end{gather}
Note that $v_1(x)$ and $v_2(x)$ vary in space through their dependence on $v(x)$ (see Eq.~\ref{eq:v1v2}).
Eliminating $p(x)$, we obtain a finite $\mathrm{v}_n(x)$ controlled by an activity gradient. The topological constraint gives $\rho(x)=\partial_x\mathrm{v}_n(x)/(2\pi)$, which in analogy with Gauss' law demonstrates that a transverse phase gradient, here set up by a gradient in activity, acts as a local ``electric field''. So we have
\begin{gather}
	p(x)=-\dfrac{\nu\kappa}{2D_R}\dfrac{\partial_x\left[\vphantom{a^{1/2}_{2/3}}v_1(x)n_0(x)\right]}{\left[\nu\kappa-\dfrac{v(x)v_2(x)}{2D_R}\right]}\ ,\label{eq:px}\\
	\rho(x)=\dfrac{1}{8\pi D_R}\partial_x\left\{\dfrac{v(x)\,\partial_x\left[\vphantom{a^{1/2}_{2/3}}v_1(x)n_0(x)\right]}{n_0(x)\left[\nu\kappa-\dfrac{v(x)v_2(x)}{2D_R}\right]}\right\}\ .\label{eq:rhox}
\end{gather}
One can check that $\rho(x)$ does not depend on the sign of $v(x)$ whereas $p(x)$ does, as expected. The denominators in the expressions above don't vanish as we are below the defect ordering transition. The defect charge and number density for two different activity profiles are plotted in Fig.~\ref{fig:interface}, along with a schematic showing the polarization of defects at the active-passive interface. Hence an active-passive interface is both charge and orientationally polarized. As seen in Fig.~\ref{fig:interface}, there is an excess of $+1/2$ disclinations on the passive side of such an interface and charge balanced excess of $-1/2$ disclinations on the active side. This can be understood by recalling that the $+1/2$ defects being motile tend to accumulate where they move slowly. Similarly, an active region flanked by passive regions on either side develops a net negative topological charge concentrated near the interfaces \footnote{A related phenomenon is the spontaneous charging of isotropic tactoids in an active nematic \cite{genkin2018spontaneous}}. While the overall scale of $\rho(x)$ involves the defect persistence length and nematic elasticity, the length scale over which charges separate is directly governed by the width of the interface. One can similarly analyze the defect charge distribution setup by a spatial activity pattern that switches from extensile ($v<0$) to contractile ($v>0$). At the interface the activity is forced to vanish, leading to an accumulation of $+1/2$ defects bordered on either side by a compensating layer of $-1/2$ defects. The charge and polarization distribution continue to be given by Eqs.~\ref{eq:rhox} and \ref{eq:px}, respectively, independent of any spatial variations in the sign of activity.

\section{Discussion}
\label{sec:discussion}
Topological defects play a foundational role in characterizing ordered media, being fingerprints of broken symmetry. In active nematics, they acquire additional dynamical character due to the breakdown of detailed balance. By emphasizing the dominant role of defects as drivers of flow, we have developed a detailed hydrodynamic theory of active defects to capture the various dynamical states of a noisy 2D active nematic on a substrate. This approach allows us to analytically treat both active turbulence, and defect ordering at higher activity. Our results on spatiotemporal defect chaos in the active turbulent regime are consistent with previous numerical work \cite{giomi2015geometry,hemingway2016correlation}, and provide a tractable starting point to address the large scale flow signatures of a strongly interacting defect gas. At high enough activity, torques acting on the $+1/2$ disclinations become strong enough to collectively align the moving defects into a spontaneously flowing liquid. Our analysis identifies a definite physical mechanism that drives the defect ordering transition and explains the underlying reason for \emph{polar} defect order. Finally, extending our treatment to handle spatially inhomogeneous activity, we demonstrate that activity gradients can act as ``electric fields'' (see also Appendix~\ref{app:efield}) that can be used to corral defects and segregate them. Understanding situations where activity can be spatially or temporally manipulated is the first step to controlling and patterning structure, along with facilitating targeted transport in active matter.

\begin{figure}[]
	\centering
	\includegraphics[width=0.43\textwidth]{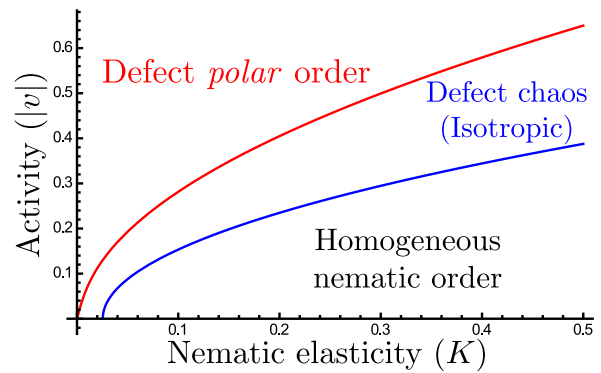}
	\caption{Phase diagram as a function of activity ($|v|$) and nematic elasticity ($K$). Here we have fixed $\nu=\pi$ and $\gamma=1$ and taken $T=T_0+T_1v^2$, and $D_R=D_R^0+D_R^1v^2$, with $T_0/T_1=10^{-2}$ and $D_R^0/D_R^1=10$ in units where $v_0^2=D_R^0T_0/\gamma=0.1$ sets the scale of $|v|^2$. The overall topology of this phase diagram does not change for other values of parameters, with the sole exception of $T_0=0,\,T_1\neq 0$ in which case the homogeneous nematic phase extends all the way down to $K=0$.}
	\label{fig:phase_dgm_new}
\end{figure}

The phase transitions demarcating the different states of defect organization are depicted in Fig.~\ref{fig:phasedgm} through a phase diagram constructed in terms of the parameters of our theoretical model. To make contact with possible experimental realizations of our predictions, it is useful to cast the phase boundaries in terms of variables that are easier to control in experiments. Both translational noise as captured by our effective temperature $T$ and rotational noise captured by $D_R$ are dominantly of nonthermal origin and \emph{a priori} unknown. Furthermore, they will generally be controlled by active processes, which renders them dependent on the system's activity. To construct a phase diagram in terms of two independent and in principle theoretically accessible axes, we assume a generic dependence of the effective temperature and the rotational diffusion on activity as $T=T_0+T_1v^2$ and $D_R=D_R^0+D_R^1v^2$, with $T_0$ and $D_R^0$ as passive contributions, and construct a phase diagram in terms of the nematic stiffness $K$ and activity $|v|$. The resulting phase diagram is shown in Fig.~\ref{fig:phase_dgm_new}. In biofilament-motor complex suspensions \cite{sanchez2012spontaneous,kumar2018tunable}, the passive elasticity of the nematic, is primarily controlled by filament length and density, whereas activity can be tuned by changing ATP and possibly motor concentration. Of course activity will also affect the nematic elasticity~\cite{kumar2018tunable}, but this can be considered a higher order effect. A different realization of an active nematic involves perfusing a biocompatible molecular liquid crystal with bacteria \cite{Zhou2014}. Here the elasticity of the liquid crystal medium is known and well-controlled, while the activity can be tuned by regulating the bacterial concentration. We look forward to experiments that will direct tests of our results.

Our work opens up several important future directions and leaves open challenges to be resolved. Arguably the most obvious one is the lack of nematic defect order in our model. As discussed in Sec.~\ref{subsec:defectorder} and sketched in Fig.~\ref{fig:stability}, the absence of nematic defect order is not the result of the coarse-graining scheme or closure approximations used. It instead is a basic feature of the structure of the dynamics of defect as quasi-particles that we use as our starting point. These equations were derived perturbatively in Ref.~\cite{shankar2018defect} and the resulting active torques only favor an effective polar alignment of $+1/2$ disclinations in an unbound defect gas. One may then ask whether a torque that favors nematic alignment of the $+1/2$ defects may be obtained at higher order in activity. We believe the following arguments demonstrate that this is not possible in the context of a dry nematic. To understand this, we first stress that the mechanism for defect ordering described here relies on two ingredients: the balance of elastic forces and propulsive fluxes to ensure $\b{j}_{\rho}\approx\b{0}$ and prevent charge segregation, and the active torques that favor local buildup of polar order in the presence of an elastic distortion.
Any active torque with nematic symmetry that were to arise in a non-perturbative treatment will have to involve an alignment 
 interaction between the $+1/2$ defect polarization and the local phase gradient ($\b{v}_n$) as these are the only vector fields in the problem. Such a torque cannot, however, support pure nematic defect order on large scales because many-body screening will force $\b{v}_n$ to decay to zero, making it impossible to balance a finite elastic distortion against an active charge current involving the nematic order parameter of defects. This seems to exclude the possibility of purely nematic defect ordering in an active nematic on a substrate. One possible way out is to include hydrodynamic interactions and momentum conserving flow. Genuine fluid flow can provide nonlocal alignment interactions in addition to new terms controlling charge transport. Solvent flow is expected to be important, at least partly, in experiments involving micotubule-kinesin suspensions, where nematic ordering of defects has been previously reported \cite{decamp2015orientational}, and is also known to stabilize defect lattices in simulations \cite{doostmohammadi2016stabilization}. Extending our framework to include viscous flow and account for nematic defect ordering is a significant challenge for the future.

Another direction for future research is to use our framework to quantify design principles for engineering active microfluidic devices. Controlling flows and defect organization through confinement \cite{keber2014topology,ellis2018curvature,guillamat2017taming,opathalage2019self} invites future investigations into the nontrivial role of boundaries and curvature in active matter. Our work on defect segregation through activity gradients poses new avenues for exploration in the context of spatio-temporal control of activity. Furthermore, in the spirit of metamaterial design, controlling flow and patterning structures in active matter is essential to use active devices for applications. In this regard, marrying optimal transport with active fluid hydrodynamics is the next step towards achieving this goal.

\acknowledgments
We thank Mark Bowick and Sriram Ramaswamy for illuminating discussions. We also acknowledge useful discussions with Juan de Pablo, Rui Zhang and Margaret Gardel. This work was supported by the National Science Foundation through the award DMR-1609208 (M.C.M and S.S.), at KITP under Grant No. PHY-1748958 and at the Aspen Center for Physics under grant PHY-1607611. S.S.~would like to thank both the KITP and the Aspen Center for Physics for their support and hospitality during the completion of some of this work.

\appendix
\section{Derivation of active defect hydrodynamics}
\label{app:calc}
We briefly recapitulate the order parameter based description of active nematics to both make contact with conventional modeling strategies and also remind ourselves of the various parameters in the model. The continuum nematodynamic equations are just as for the order parameter $\b{Q}$ of a passive 2D liquid crystal \cite{beris1994thermodynamics}, given by
\begin{equation}
	\partial_t\b{Q}+\b{u}\cdot\bm\del\b{Q}+\left[\vec{\Omega},\b{Q}\right]=\lambda\left(\bm\del\b{u}\right)_{ST}+\dfrac{1}{\gamma}\b{H}\ ,\label{eq:Qeq}
\end{equation}
with $\b{u}$ the flow velocity, $2\Omega_{\mu\nu}=\del_{\mu}u_{\nu}-\del_{\nu}u_{\mu}$ the vorticity tensor and $\gamma$ the rotational viscosity. The flow alignment term involves the coupling $\lambda$ and the deviatoric strain rate tensor $(\del_{\mu}u_{\nu})_{ST}=\del_{\mu}u_{\nu}+\del_{\nu}u_{\mu}-\delta_{\mu\nu}(\bm\del\cdot\b{u})$. The molecular field $\b{H}=b(1-S^2)\b{Q}+K\del^2\b{Q}$ with $b>0$ controls the mean-field isotropic to nematic transition at equilibrium and $K$ is the Frank elastic constant.

In the presence of a frictional substrate and activity, we supplement this with a force balance equation for the flow velocity, involving active stresses, $-\Gamma\b{u}+\bm\del\cdot\vec{\sigma}^a=\b{0}$. We have assumed that friction with the substrate $\Gamma$ screens the flow on large scales and have retained only the dominant active stress, $\vec{\sigma}^a=\alpha\b{Q}$, neglecting all passive elastic contributions. The activity $\alpha$ corresponds to the average force dipole exerted by the microscopic apolar active units on the surrounding fluid \cite{ramaswamy2010mechanics,marchetti2013hydrodynamics}, with $\alpha>0$ for contractile systems and $\alpha<0$ for extensile systems. For an isolated $+1/2$ disclination positioned at $\b{r}_i^+$, the active flow generated at the core of the defect is $\b{u}(\b{r}_i^+)=v\b{e}_i$, where $v=\alpha/(a\Gamma)$ relates the defect motility to the activity and $\b{e}_i=a\bm\del\cdot\b{Q}(\b{r}_i^+)$ is the defect orientation as mentioned in the main text. Below we shall derive the hydrodynamic description of active defects.

\subsection{Phase gradient dynamics}
The topological constraint (Eq.~\ref{eq:topo}) relates the phase gradient $\b{v}_n$ to the defect charge density through $\hat{\b{z}}\cdot\left(\bm\del\times\b{v}_n\right)=2\pi\rho$.
As defects are created or annihilated only in pairs, the charge density $\rho$ is locally conserved and we have a continuity equation for the dynamics of $\rho$, as noted in Eq.~\ref{eq:rhoeq}
\begin{equation}
	\partial_t\rho+\bm\del\cdot\b{j}_{\rho}=0\ .\label{eq:ncont}
\end{equation}
Note that this expression is valid at the fluctuating level in the Stratanovich convention. As the phase gradient $\b{v}_n$ is single valued, we can commute derivatives on it. Writing $2\pi\rho=\hat{\b{z}}\cdot\left(\bm\del\times\b{v}_n\right)$ and commuting the time and space derivatives, Eq.~\ref{eq:ncont} gives
\begin{gather}
	\bm\del\times\left(\partial_t\b{v}_n-2\pi\vec{\e}\cdot\b{j}_{\rho}\right)=\bm 0\nonumber\\
	\implies\partial_t\b{v}_n-2\pi\vec{\e}\cdot\b{j}_{\rho}=\del\Pi\ ,\label{eq:dvsdt1}
\end{gather}
where $\vec{\e}$ is the 2D Levi-Civita tensor and $\Pi$ is a smooth scalar function free of singularities. Away from defects $\b{j}_{\rho}=\b{0}$, and a smooth director fluctuation then obeys
\begin{equation}
	\partial_t\theta=\dfrac{K}{\gamma}\del^2\theta+\sqrt{\dfrac{2T}{\gamma}}f_{\theta}\ ,\label{eq:XY}
\end{equation}
on large scales, where $f_{\theta}$ is unit white Gaussian noise and $T$ is the noise strength that functions as an effective temperature. All nonlinearities affecting the smooth phase fluctuations in the ordered 2D active nematic are known to be perturbatively irrelevant on large enough length scales \cite{mishra2010dynamic,shankar2018low}. Comparing Eq.~\ref{eq:XY} and Eq.~\ref{eq:dvsdt1}, we set $\Pi=(K/\gamma)\bm\del\cdot\b{v}_n+\sqrt{2T/\gamma}f_{\theta}$. Putting it all together, we then have
\begin{equation}
	\partial_t\b{v}_n=2\pi\,\vec{\e}\cdot\b{j}_{\rho}+\dfrac{K}{\gamma}\bm\del(\bm\del\cdot\b{v}_n)+\sqrt{\dfrac{2T}{\gamma}}\bm\del f_{\theta}\ .\label{eq:dvsdt}
\end{equation}
Upon setting $\kappa=K/\gamma$ and neglecting the noise in a mean-field description, we obtain Eq.~\ref{eq:vseq} in the main text.

\subsection{Current constitutive equation}
The motion of a $\pm1/2$ disclination in an active nematic was derived in Ref.~\cite{shankar2018defect} to be
\begin{subequations}
\begin{align}
	\dot{\b{r}}^+_i&=v\b{e}_i+\pi\mu K\,\vec{\e}\cdot\b{v}_n+\sqrt{2\mu T}\vec{\xi}_i(t)\ ,\label{eq:eom+}\\
	\dot{\b{r}}^-_i&=-\pi\mu K\,\vec{\e}\cdot\b{v}_n+\sqrt{2\mu T}\vec{\xi}_i(t)\ ,\label{eq:eom-}
\end{align}\label{eq:eom}
\end{subequations}
where $\mu\propto\gamma^{-1}$ is the defect mobility and $\vec{\xi}_i(t)$ is unit white noise. We emphasize that $v\propto\alpha$ can be of either sign, with extensile and contractile systems propelling the $+1/2$ defect in opposite directions \cite{giomi2013defect}, while the $-1/2$ defect remains diffusive and non-motile.
The particular form of the Magnus-like force in Eq.~\ref{eq:eom} was chosen to recover the passive Coulomb interaction between bound defects. In the absence of free unbound defects, the phase gradient at any point due to bound defect pairs with charge $q_i=\pm1/2$ is
\begin{equation}
	\b{v}_n(\b{r})=-\vec{\e}\cdot\bm\del\sum_iq_i\ln\left|\dfrac{\b{r}-\b{r}_i}{a}\right|\ ,\label{eq:vsbound}
\end{equation}
where $\b{r}_i$ is the position of the $i$th defect and $a$ is the defect core size that provides a microscopic cutoff. Using Eq.~\ref{eq:vsbound} in Eq.~\ref{eq:eom}, one can check that the correct form of the passive elastic force is obtained when considering just bound defect pairs. In the presence of unbound defects, Eq.~\ref{eq:vsbound} is no longer applicable and we instead have to use Eq.~\ref{eq:dvsdt} to obtain the phase gradient.

Using the defect equations of motion (Eq.~\ref{eq:eom}) in the definition of the fluctuating current ($\b{j}_{\pm}=\sum_i\dot{\b{r}}^{\pm}_i\delta(\b{r}-\b{r}_i^{\pm})$), we obtain
\begin{align}
	\b{j}_{+}&=v\b{p}+\pi\mu K\,n_+\vec{\e}\cdot\b{v}_n-\mu T\bm\del n_++\sqrt{2\mu Tn_+}\,\vec{\xi}_+\ ,\\
	\b{j}_-&=-\pi\mu K\,n_-\vec{\e}\cdot\b{v}_n-\mu T\bm\del n_-+\sqrt{2\mu Tn_-}\,\vec{\xi}_-\ ,
\end{align}
where $\vec{\xi}_{\pm}(\b{r},t)$ are unit space-time white noises. For spatially varying $n_{\pm}$, the multiplicative noise must be interpreted in Ito style. Writing $D_0=\mu T$, replacing $\pi\mu K$ by $\nu\kappa$, and neglecting noise at the mean-field level, we obtain Eqs.~\ref{eq:jp} and~\ref{eq:jm}. The corresponding fluctuating expressions for $\b{j}_{\rho}$ and $\b{j}_{n}$ can then be trivially obtained.
\subsection{Polarization dynamics}
\label{app:poldyn}
The orientational dynamics of $\b{e}_i$ including active torques was derived in Ref.~\cite{shankar2018defect} to be
\begin{align}
	\dot{\b{e}}_i&=-\dfrac{5\pi\mu\gamma}{8}\left[\pi\mu K\,|\b{v}_n|^2+v\,\hat{\b{z}}\cdot\left(\b{e}_i\times\b{v}_n\right)\right]\b{e}_i\nonumber\\
	&\quad-v\dfrac{\pi\mu\gamma}{8}\left(\b{e}_i\cdot\b{v}_n\right)\vec{\e}\cdot\b{e}_i-\dfrac{2K}{\gamma}\left(\bm\del\cdot\b{v}_n\right)\vec{\e}\cdot\b{e}_i\nonumber\\
	&\quad-\sqrt{2D_R}\,\vec{\e}\cdot\b{e}_i\,\eta_i(t)+\vec{\nu}_i(t)\ .\label{eq:edot}
\end{align}
This is written here in a form that is appropriate for treating unbound defects. Unpackaging the various terms, the top line of Eq.~\ref{eq:edot} involves both passive and active terms that relax or enhance $|\b{e}_i|$, which is a measure of local nematic distortion. The second line of Eq.~\ref{eq:edot} includes orientational torques that leave $|\b{e}_i|$ fixed. The first of these is the important active torque that causes the $+1/2$ defect orientation to self-align with the local phase gradient. The second is a passive elastic torque that causes the $+1/2$ defect to reorient in a splayed phase gradient. This can be easily obtained by noting that, away from defects, an external phase gradient causes the nematic director to precess at a rate given by $\bm\del\cdot\b{v}_n$ (from Eq.~\ref{eq:XY}). Following the analysis of Ref.~\cite{shankar2018defect}, this generates the required elastic torque in Eq.~\ref{eq:edot}. As can be seen, this elastic torque is subdominant to the active one, as it only arises for $\bm\del\cdot\b{v}_n\neq 0$. Similar elastic torques have been obtained by different means previously as well \cite{keber2014topology,vromans2016orientational,tang2017orientation}. The final two terms in the last line of Eq.~\ref{eq:edot} are noise, with $D_R$ as the rotational diffusion constant, $\eta_i(t)$ as unit white Gaussian noise, and the longitudinal noise $\vec{\nu}_i(t)$ is Gaussian with zero mean and correlations
\begin{equation}
	\langle\vec{\nu}_i(t)\vec{\nu}_j(t')\rangle=T\dfrac{5\pi^2}{4}\mu^2\gamma|\b{v}_n|^2\,\b{1}\,\delta_{ij}\delta(t-t')\ .
\end{equation}
The form of the noise correlator is obtained by using essentially a fluctuation-dissipation like relation in the limit $v=0$.
For simplicity we use the same effective noise strength $\sim T$ here as other choices don't change the results in any qualitative way (also see Ref.~\cite{shankar2018defect}).
Coarse-graining Eq.~\ref{eq:edot}, we obtain a fluctuating hydrodynamic equation for the defect polarization density $\b{p}$,
\begin{widetext}
\begin{align}
	\partial_tp_{\alpha}+v\partial_{\beta}M_{\alpha\beta}+\pi\mu K\partial_{\beta}\left[\e_{\beta\gamma}\mathrm{v}_{n\gamma}p_{\alpha}\right]&=-\left[D_R+\dfrac{5\pi^2\mu^2\gamma K}{8}|\b{v}_n|^2\right]p_{\alpha}-v\dfrac{\pi\mu\gamma}{2}M_{\alpha\beta}\e_{\beta\gamma}\mathrm{v}_{n\gamma}\nonumber\\
	&\quad-v\dfrac{\pi\mu\gamma}{8}(\e_{\alpha\beta}\mathrm{v}_{n\beta})\tr(\b{M})+\mu T\del^2p_{\alpha}-2\e_{\alpha\beta}p_{\beta}(\del\cdot\b{v}_n)+\sqrt{D_R\tr(\b{M})}\Lambda_{\alpha}\ .\label{eq:Apeq}
\end{align}
\end{widetext}
This equation is not closed as it involves the second order tensorial moment $\b{M}(\b{r},t)=\sum_i\b{e}_i(t)\b{e}_i(t)\delta(\b{r}-\b{r}_i^+(t))$. The Gaussian noise $\vec{\Lambda}$ is spatio-temporally unit white. We have also neglected subdominant terms in the noise involving $n_+|\b{v}_n|^2$ and the traceless part of $\b{M}$, along with $\mathcal{O}(\del^2)$ contributions. These do not affect the results presented.

To close the moment heirarchy, we use a Ginzburg-Landau ansatz taking both $\b{p}$ and $\b{v}_n$ to be equally small and slowly varying. In addition, we set $\b{M}=(m/2)\b{1}$ and disregard all higher anisotropic moments of the orientation that decay on short time scales $\sim D_R^{-1}$, neglecting any nematic ordering of the $+1/2$ defects for now. As the active torques only generate alignment between $\b{p}$ and $\b{v}_n$, within our model it is impossible to obtain nematic order of defects directy from the isotropic state. The last thing left to do is then determine $\tr(\b{M})=m$ in terms of the variables retained. At lowest order, without noise, we find
\begin{gather}
	m=\dfrac{2T}{K}f\left(\zeta\right)n_++\mathcal{O}\left(\b{v}_n^2,\del^2\right)\ ,\\
	\zeta=\dfrac{v^2\gamma T}{D_RK^2}\ ,
\end{gather}
where $\zeta$ is a nondimensional activity. The positive dimensionless function $f(\zeta)=1+4\zeta+\mathcal{O}(\zeta^2)$ involves the leading $\zeta$ correction arising from eliminating the third order moment [$\langle\sum_i\b{e}_i|\b{e}_i|^2\delta(\b{r}-\b{r}_i^+)\rangle\simeq -v\pi\mu\gamma(8T^2/D_RK^2)n_+\,\vec{\e}\cdot\b{v}_n$]. This completes the derivation of defect hydrodynamics.

\section{Defect alignment and electric field analogy}
\label{app:efield}

\begin{figure}[]
	\centering
	\includegraphics[width=0.5\textwidth]{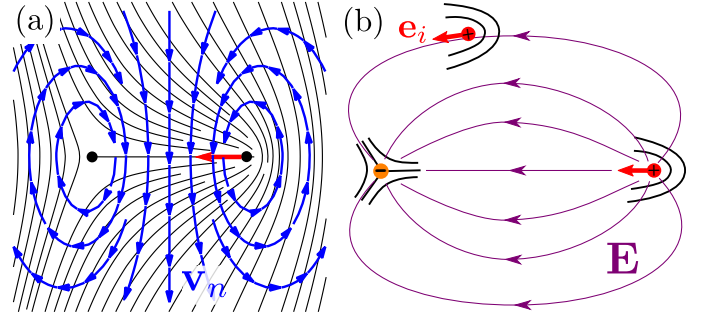}
	\caption{(a) The elastic distortion created by a neutral pair of $\pm1/2$ defects at a finite separation. The director is marked in black and the phase gradient $\b{v}_n$ is in blue. (b) A schematic with the neutral defect pair in the same configuration as in (a). The analogue electric field $\b{E}=\vec{\b{e}}\cdot\b{v}_n$ shown in purple lines traces out the familiar electric field lines between a pair of opposite sign electric charges. For extensile activity ($v<0$), a test $+1/2$ disclination shown above the neutral pair aligns its polarization $\b{e}_i$ with the ``electric field'' generated by the defect pair.  This configuration shown here is the same as shown in Fig.~\ref{fig:stability}, but now reinterpreted in terms of the analogue electric field $\b{E}$. The contractile case works similarly, with some of the signs switched.}
	\label{fig:app}
\end{figure}
In this Appendix we provide further details on the alignment torques acting on the $+1/2$ disclinations, with an intuitive interpretation in terms of an electrostatic analogy. The full equation for the angular dynamics of the $+1/2$ defect orientation is given in Appendix~\ref{app:poldyn}. To develop the electrostatic analogy, we only consider the angular dynamics of the $+1/2$ defect and neglect noise and passive elastic torques. Writing $\b{e}_i=|\b{e}_i|(\cos\psi_i,\sin\psi_i)$ for an individual defect polarization, in the presence of a finite phase distortion $\b{v}_n$ the active torques generate an angular velocity  given by
\begin{equation}
	\dot{\psi}_i=v\dfrac{\nu}{8}\left(\b{e}_i\cdot\b{v}_n\right)=-\dfrac{\nu}{8}\partial_{\psi_i}\mathcal{H}\ .
\end{equation}
For the last equality, we have recast the active alignment torque as deriving from an effective alignment ``energy'' $\mathcal{H}=v\sum_i\hat{\b{z}}\cdot(\b{e}_i\times\b{v}_n)$. This is simply a formal rewriting of the dynamical torque with no real energetic underpinning. Now, from the topological charge constraint $\hat{\b{z}}\cdot(\bm\del\times\b{v}_n)=2\pi\rho$ (Eq.~\ref{eq:topo}), we identify an ``electric field'' $\b{E}=\vec{\e}\cdot\b{v}_n$. This allows us to rewrite  the charge constraint in the form of Gauss' law,
\begin{equation}
	\bm\del\cdot\b{E}=2\pi\rho\ .
\end{equation}
This identification is the same as in the Maxwell analogy for superfluid dynamics \cite{ambegaokar1980dynamics}. As discussed in Sec.~\ref{sec:inhomo}, an activity gradient can enforce a locally nonvanishing $\b{v}_n$, which in turn acts as an ``electric field'' through the above identification.

We can then rewrite the effective alignment interaction $\mathcal{H}$ in terms of the electric field $\b{E}$ as
\begin{equation}
	\mathcal{H}=v\sum_i\b{e}_i\cdot\b{E}\ .\label{eq:dipole}
\end{equation}
It is then evident that the active alignment of a neutral defect pair takes exactly the form of the alignment of an electric dipole with dipole moment $-v\b{e}_i$ in an electric field $\b{E}$. This allows us to easily and intuitively interpret the consequences of the active torque. As shown in the schematic in Fig.~\ref{fig:stability} and in more detail  in Fig.~\ref{fig:app}, a neutral pair of defects generates an elastic distortion or phase gradient $\b{v}_n$  plotted as blue lines in Fig.~\ref{fig:app}a. Conversely, one can reinterpret the same picture in terms of $\b{E}$, as shown in Fig.~\ref{fig:app}b. Here the ``electric field'' lines are shown in purple and are orthogonal to the local phase gradient. As stated in Eq.~\ref{eq:dipole}, the active torque can effectively be seen as akin to a dipole alignment interaction. In the extensile case ($v<0$), the effective dipole moment is $|v|\b{e}_i$ and preferentially aligns with the local electric field. A similar argument works for the contractile case as well with the signs flipped. In both cases, the effective alignment between the $+1/2$ defects is \emph{polar}. In short, Fig.~\ref{fig:app} provides a different interpretation of the alignment  induced by active torques (as sketched in Fig.~\ref{fig:stability}) in terms of an electrostatic analogy.


\begin{thebibliography}{10}

\bibitem{wensink2012meso}
Henricus~H Wensink, J{\"o}rn Dunkel, Sebastian Heidenreich, Knut Drescher,
  Raymond~E Goldstein, Hartmut L{\"o}wen, and Julia~M Yeomans.
\newblock Meso-scale turbulence in living fluids.
\newblock {\em Proceedings of the National Academy of Sciences},
  109(36):14308--14313, 2012.

\bibitem{saw2017topological}
Thuan~Beng Saw, Amin Doostmohammadi, Vincent Nier, Leyla Kocgozlu, Sumesh
  Thampi, Yusuke Toyama, Philippe Marcq, Chwee~Teck Lim, Julia~M Yeomans, and
  Benoit Ladoux.
\newblock Topological defects in epithelia govern cell death and extrusion.
\newblock {\em Nature}, 544(7649):212, 2017.

\bibitem{kawaguchi2017topological}
Kyogo Kawaguchi, Ryoichiro Kageyama, and Masaki Sano.
\newblock Topological defects control collective dynamics in neural progenitor
  cell cultures.
\newblock {\em Nature}, 545(7654):327, 2017.

\bibitem{blanch2018turbulent}
C~Blanch-Mercader, V~Yashunsky, S~Garcia, G~Duclos, L~Giomi, and P~Silberzan.
\newblock Turbulent dynamics of epithelial cell cultures.
\newblock {\em Physical review letters}, 120(20):208101, 2018.

\bibitem{li2019data}
He~Li, Xia-qing Shi, Mingji Huang, Xiao Chen, Minfeng Xiao, Chenli Liu, Hugues
  Chat{\'e}, and HP~Zhang.
\newblock Data-driven quantitative modeling of bacterial active nematics.
\newblock {\em Proceedings of the National Academy of Sciences},
  116(3):777--785, 2019.

\bibitem{ramaswamy2010mechanics}
Sriram Ramaswamy.
\newblock The mechanics and statistics of active matter.
\newblock {\em Annu. Rev. Condens. Matter Phys.}, 1(1):323--345, 2010.

\bibitem{marchetti2013hydrodynamics}
M~Cristina Marchetti, Jean-Fran{\c{c}}ois Joanny, Sriram Ramaswamy,
  Tanniemola~B Liverpool, Jacques Prost, Madan Rao, and R~Aditi Simha.
\newblock Hydrodynamics of soft active matter.
\newblock {\em Reviews of Modern Physics}, 85(3):1143, 2013.

\bibitem{ramaswamy2003active}
Sriram Ramaswamy, R~Aditi Simha, and John Toner.
\newblock Active nematics on a substrate: Giant number fluctuations and
  long-time tails.
\newblock {\em EPL (Europhysics Letters)}, 62(2):196, 2003.

\bibitem{doostmohammadi2018active}
Amin Doostmohammadi, Jordi Ign{\'e}s-Mullol, Julia~M Yeomans, and Francesc
  Sagu{\'e}s.
\newblock Active nematics.
\newblock {\em Nature communications}, 9(1):3246, 2018.

\bibitem{sanchez2012spontaneous}
Tim Sanchez, Daniel~TN Chen, Stephen~J DeCamp, Michael Heymann, and Zvonimir
  Dogic.
\newblock Spontaneous motion in hierarchically assembled active matter.
\newblock {\em Nature}, 491(7424):431, 2012.

\bibitem{keber2014topology}
Felix~C Keber, Etienne Loiseau, Tim Sanchez, Stephen~J DeCamp, Luca Giomi,
  Mark~J Bowick, M~Cristina Marchetti, Zvonimir Dogic, and Andreas~R Bausch.
\newblock Topology and dynamics of active nematic vesicles.
\newblock {\em Science}, 345(6201):1135--1139, 2014.

\bibitem{ellis2018curvature}
Perry~W Ellis, Daniel~JG Pearce, Ya-Wen Chang, Guillermo Goldsztein, Luca
  Giomi, and Alberto Fernandez-Nieves.
\newblock Curvature-induced defect unbinding and dynamics in active nematic
  toroids.
\newblock {\em Nature Physics}, 14(1):85, 2018.

\bibitem{kumar2018tunable}
Nitin Kumar, Rui Zhang, Juan~J de~Pablo, and Margaret~L Gardel.
\newblock Tunable structure and dynamics of active liquid crystals.
\newblock {\em Science advances}, 4(10):eaat7779, 2018.

\bibitem{narayan2007long}
Vijay Narayan, Sriram Ramaswamy, and Narayanan Menon.
\newblock Long-lived giant number fluctuations in a swarming granular nematic.
\newblock {\em Science}, 317(5834):105--108, 2007.

\bibitem{Zhou2014}
Shuang Zhou, Andrey Sokolov, Oleg~D. Lavrentovich, and Igor~S. Aranson.
\newblock Living liquid crystals.
\newblock {\em Proceedings of the National Academy of Sciences},
  111(4):1265--1270, 2014.

\bibitem{nishiguchi2017long}
Daiki Nishiguchi, Ken~H Nagai, Hugues Chat{\'e}, and Masaki Sano.
\newblock Long-range nematic order and anomalous fluctuations in suspensions of
  swimming filamentous bacteria.
\newblock {\em Physical Review E}, 95(2):020601, 2017.

\bibitem{duclos2017topological}
Guillaume Duclos, Christoph Erlenk{\"a}mper, Jean-Fran{\c{c}}ois Joanny, and
  Pascal Silberzan.
\newblock Topological defects in confined populations of spindle-shaped cells.
\newblock {\em Nature Physics}, 13(1):58, 2017.

\bibitem{voituriez2005spontaneous}
R~Voituriez, Jean-Fran{\c{c}}ois Joanny, and Jacques Prost.
\newblock Spontaneous flow transition in active polar gels.
\newblock {\em EPL (Europhysics Letters)}, 70(3):404, 2005.

\bibitem{duclos2018spontaneous}
G~Duclos, C~Blanch-Mercader, V~Yashunsky, G~Salbreux, J-F Joanny, J~Prost, and
  P~Silberzan.
\newblock Spontaneous shear flow in confined cellular nematics.
\newblock {\em Nature physics}, 14(7):728, 2018.

\bibitem{thampi2013velocity}
Sumesh~P Thampi, Ramin Golestanian, and Julia~M Yeomans.
\newblock Velocity correlations in an active nematic.
\newblock {\em Physical review letters}, 111(11):118101, 2013.

\bibitem{giomi2013defect}
Luca Giomi, Mark~J Bowick, Xu~Ma, and M~Cristina Marchetti.
\newblock Defect annihilation and proliferation in active nematics.
\newblock {\em Physical review letters}, 110(22):228101, 2013.

\bibitem{shankar2018defect}
Suraj Shankar, Sriram Ramaswamy, M~Cristina Marchetti, and Mark~J Bowick.
\newblock Defect unbinding in active nematics.
\newblock {\em Physical review letters}, 121(10):108002, 2018.

\bibitem{shi2013topological}
Xia-qing Shi and Yu-qiang Ma.
\newblock Topological structure dynamics revealing collective evolution in
  active nematics.
\newblock {\em Nature communications}, 4:3013, 2013.

\bibitem{thampi2014instabilities}
Sumesh~P Thampi, Ramin Golestanian, and Julia~M Yeomans.
\newblock Instabilities and topological defects in active nematics.
\newblock {\em EPL (Europhysics Letters)}, 105(1):18001, 2014.

\bibitem{gao2015multiscale}
Tong Gao, Robert Blackwell, Matthew~A Glaser, Meredith~D Betterton, and
  Michael~J Shelley.
\newblock Multiscale polar theory of microtubule and motor-protein assemblies.
\newblock {\em Physical review letters}, 114(4):048101, 2015.

\bibitem{giomi2015geometry}
Luca Giomi.
\newblock Geometry and topology of turbulence in active nematics.
\newblock {\em Physical Review X}, 5(3):031003, 2015.

\bibitem{hemingway2016correlation}
Ewan~J Hemingway, Prashant Mishra, M~Cristina Marchetti, and Suzanne~M
  Fielding.
\newblock Correlation lengths in hydrodynamic models of active nematics.
\newblock {\em Soft Matter}, 12(38):7943--7952, 2016.

\bibitem{alert2019universal}
Ricard Alert, Jean-Fran{\c{c}}ois Joanny, and Jaume Casademunt.
\newblock Universal scaling of active nematic turbulence.
\newblock {\em arXiv preprint arXiv:1906.04757}, 2019.

\bibitem{guillamat2017taming}
Pau Guillamat, Jordi Ign{\'e}s-Mullol, and Francesc Sagu{\'e}s.
\newblock Taming active turbulence with patterned soft interfaces.
\newblock {\em Nature communications}, 8(1):564, 2017.

\bibitem{lemma2018statistical}
Linnea~M Lemma, Stephen~J Decamp, Zhihong You, Luca Giomi, and Zvonimir Dogic.
\newblock Statistical properties of autonomous flows in 2d active nematics.
\newblock {\em arXiv preprint arXiv:1809.06938}, 2018.

\bibitem{martinez2019selection}
Berta Mart{\'\i}nez-Prat, Jordi Ign{\'e}s-Mullol, Jaume Casademunt, and
  Francesc Sagu{\'e}s.
\newblock Selection mechanism at the onset of active turbulence.
\newblock {\em Nature Physics}, page~1, 2019.

\bibitem{decamp2015orientational}
Stephen~J DeCamp, Gabriel~S Redner, Aparna Baskaran, Michael~F Hagan, and
  Zvonimir Dogic.
\newblock Orientational order of motile defects in active nematics.
\newblock {\em Nature materials}, 14(11):1110, 2015.

\bibitem{putzig2016instabilities}
Elias Putzig, Gabriel~S Redner, Arvind Baskaran, and Aparna Baskaran.
\newblock Instabilities, defects, and defect ordering in an overdamped active
  nematic.
\newblock {\em Soft Matter}, 12(17):3854--3859, 2016.

\bibitem{srivastava2016negative}
Pragya Srivastava, Prashant Mishra, and M~Cristina Marchetti.
\newblock Negative stiffness and modulated states in active nematics.
\newblock {\em Soft Matter}, 12(39):8214--8225, 2016.

\bibitem{patelli2019understanding}
Aurelio Patelli, Ilyas Djafer-Cherif, Igor~S Aranson, Eric Bertin, and Hugues
  Chat{\'e}.
\newblock Understanding dense active nematics from microscopic models.
\newblock {\em arXiv preprint arXiv:1904.12708}, 2019.

\bibitem{doostmohammadi2016stabilization}
Amin Doostmohammadi, Michael~F Adamer, Sumesh~P Thampi, and Julia~M Yeomans.
\newblock Stabilization of active matter by flow-vortex lattices and defect
  ordering.
\newblock {\em Nature communications}, 7:10557, 2016.

\bibitem{oza2016antipolar}
Anand~U Oza and J{\"o}rn Dunkel.
\newblock Antipolar ordering of topological defects in active liquid crystals.
\newblock {\em New Journal of Physics}, 18(9):093006, 2016.

\bibitem{yaman2018emergence}
Yusuf~Ilker Yaman, Esin Demir, Roman Vetter, and Askin Kocabas.
\newblock Emergence of active nematics in chaining bacterial biofilms.
\newblock {\em Nature communications}, 10(1):2285, 2019.

\bibitem{doostmohammadi2016defect}
Amin Doostmohammadi, Sumesh~P Thampi, and Julia~M Yeomans.
\newblock Defect-mediated morphologies in growing cell colonies.
\newblock {\em Physical review letters}, 117(4):048102, 2016.

\bibitem{chaikin2000principles}
Paul~M Chaikin and Tom~C Lubensky.
\newblock {\em Principles of condensed matter physics}.
\newblock Cambridge university press, 2000.

\bibitem{pismen2013dynamics}
LM~Pismen.
\newblock Dynamics of defects in an active nematic layer.
\newblock {\em Physical Review E}, 88(5):050502, 2013.

\bibitem{ambegaokar1980dynamics}
Vinay Ambegaokar, BI~Halperin, David~R Nelson, and Eric~D Siggia.
\newblock Dynamics of superfluid films.
\newblock {\em Physical Review B}, 21(5):1806, 1980.

\bibitem{volovik1980hydrodynamics}
GE~Volovik and VS~Dotsenko.
\newblock Hydrodynamics of defects in condensed media, using as examples
  vortices in rotating he ii and disclinations in a planar magnet.
\newblock {\em Soviet Phys. JETP}, 78(1):132--148, 1980.

\bibitem{marchetti1990hydrodynamics}
M~Cristina Marchetti and David~R Nelson.
\newblock Hydrodynamics of flux liquids.
\newblock {\em Physical Review B}, 42(16):9938, 1990.

\bibitem{zippelius1980dynamics}
Annette Zippelius, BI~Halperin, and David~R Nelson.
\newblock Dynamics of two-dimensional melting.
\newblock {\em Physical Review B}, 22(5):2514, 1980.

\bibitem{Note1}
Our presentation closely follows Refs.~\cite
  {ambegaokar1980dynamics,zippelius1980dynamics}.

\bibitem{Jackson1975}
J~David Jackson.
\newblock {\em Electrodynamics}.
\newblock Wiley Online Library, 1975.

\bibitem{schuppler2016boundaries}
Matthias Schuppler, Felix~C Keber, Martin Kr{\"o}ger, and Andreas~R Bausch.
\newblock Boundaries steer the contraction of active gels.
\newblock {\em Nature communications}, 7:13120, 2016.

\bibitem{ross2018controlling}
Tyler~D Ross, Heun~Jin Lee, Zijie Qu, Rachel~A Banks, Rob Phillips, and Matt
  Thomson.
\newblock Controlling organization and forces in active matter through
  optically-defined boundaries.
\newblock {\em arXiv preprint arXiv:1812.09418}, 2018.

\bibitem{cortese2018pair}
Dario Cortese, Jens Eggers, and Tanniemola~B Liverpool.
\newblock Pair creation, motion, and annihilation of topological defects in
  two-dimensional nematic liquid crystals.
\newblock {\em Physical Review E}, 97(2):022704, 2018.

\bibitem{tang2019theory}
Xingzhou Tang and Jonathan~V Selinger.
\newblock Theory of defect motion in 2d passive and active nematic liquid
  crystals.
\newblock {\em Soft matter}, 15(4):587--601, 2019.

\bibitem{lamb1993hydrodynamics}
Horace Lamb.
\newblock {\em Hydrodynamics}.
\newblock Cambridge university press, 1993.

\bibitem{peach1950forces}
M~Peach and JS~Koehler.
\newblock The forces exerted on dislocations and the stress fields produced by
  them.
\newblock {\em Physical Review}, 80(3):436, 1950.

\bibitem{kosterlitz1973ordering}
John~Michael Kosterlitz and David~James Thouless.
\newblock Ordering, metastability and phase transitions in two-dimensional
  systems.
\newblock {\em Journal of Physics C: Solid State Physics}, 6(7):1181, 1973.

\bibitem{stein1978kosterlitz}
DL~Stein.
\newblock Kosterlitz-thouless phase transitions in two-dimensional liquid
  crystals.
\newblock {\em Physical Review B}, 18(5):2397, 1978.

\bibitem{szabo2006phase}
Balint Szabo, GJ~Sz{\"o}ll{\"o}si, B~G{\"o}nci, Zs~Jur{\'a}nyi, David Selmeczi,
  and Tam{\'a}s Vicsek.
\newblock Phase transition in the collective migration of tissue cells:
  experiment and model.
\newblock {\em Physical Review E}, 74(6):061908, 2006.

\bibitem{henkes2011active}
Silke Henkes, Yaouen Fily, and M~Cristina Marchetti.
\newblock Active jamming: Self-propelled soft particles at high density.
\newblock {\em Physical Review E}, 84(4):040301, 2011.

\bibitem{weber2013long}
Christoph~A Weber, Timo Hanke, J~Deseigne, S~L{\'e}onard, Olivier Dauchot,
  Erwin Frey, and Hugues Chat{\'e}.
\newblock Long-range ordering of vibrated polar disks.
\newblock {\em Physical review letters}, 110(20):208001, 2013.

\bibitem{lam2015self}
Khanh-Dang Nguyen~Thu Lam, Michael Schindler, and Olivier Dauchot.
\newblock Self-propelled hard disks: implicit alignment and transition to
  collective motion.
\newblock {\em New Journal of Physics}, 17(11):113056, 2015.

\bibitem{abrikosov1957magnetic}
Alexei~A Abrikosov.
\newblock On the magnetic properties of superconductors of the second group.
\newblock {\em Sov. Phys. JETP}, 5:1174--1182, 1957.

\bibitem{renn1988abrikosov}
Scott~R Renn and Tom~C Lubensky.
\newblock Abrikosov dislocation lattice in a model of the
  cholesteric--to--smectic-a transition.
\newblock {\em Physical Review A}, 38(4):2132, 1988.

\bibitem{martin1988sum}
Ph~A Martin.
\newblock Sum rules in charged fluids.
\newblock {\em Reviews of Modern Physics}, 60(4):1075, 1988.

\bibitem{shankar2020}
Suraj Shankar, Sriram Ramaswamy, M~Cristina Marchetti, and Mark~J Bowick.
\newblock in preparation.

\bibitem{toner2012birth}
John Toner.
\newblock Birth, death, and flight: A theory of malthusian flocks.
\newblock {\em Physical review letters}, 108(8):088102, 2012.

\bibitem{romanczuk2016emergent}
Pawel Romanczuk, Hugues Chat{\'e}, Leiming Chen, Sandrine Ngo, and John Toner.
\newblock Emergent smectic order in simple active particle models.
\newblock {\em New Journal of Physics}, 18(6):063015, 2016.

\bibitem{adhyapak2013live}
Tapan~Chandra Adhyapak, Sriram Ramaswamy, and John Toner.
\newblock Live soap: stability, order, and fluctuations in apolar active
  smectics.
\newblock {\em Physical review letters}, 110(11):118102, 2013.

\bibitem{chen2013universality}
Leiming Chen, John Toner, et~al.
\newblock Universality for moving stripes: A hydrodynamic theory of polar
  active smectics.
\newblock {\em Physical review letters}, 111(8):088701, 2013.

\bibitem{Note2}
Though, from Eq.~\ref {eq:gamma||}, it might seem that $\delta p_{||}$ can go
  unstable at high activity when $\lambda _2\lambda _3>0$, this is an artifact
  of the low ${\protect \bf q}$ expansion. In fact, for $\varphi =\pi /2$, the
  complete dispersion relation for the amplitude mode is $i\omega
  _{||}=[\protect \mathaccentV {tilde}07E{a}_4+(2D_0+D_1)q^2+\protect \sqrt
  {(\protect \mathaccentV {tilde}07E{a}_4-D_1q^2)^2-4q^2\protect \mathaccentV
  {tilde}07E{a}_4\Lambda }]/2$, where $\protect \mathaccentV
  {tilde}07E{a}_4=2a_4D_Rp_0^2$ and $\Lambda =\lambda _2\lambda _3/a_4D_R$. For
  either sign of $\Lambda $, $\protect \mathrm {Re}(i\omega _{||})>0$ for all
  $q$, confirming that the only possible instability is the orientational one
  obtained in Eq.~\ref {eq:splay}.

\bibitem{tailleur2008statistical}
J~Tailleur and ME~Cates.
\newblock Statistical mechanics of interacting run-and-tumble bacteria.
\newblock {\em Physical review letters}, 100(21):218103, 2008.

\bibitem{fily2012athermal}
Yaouen Fily and M~Cristina Marchetti.
\newblock Athermal phase separation of self-propelled particles with no
  alignment.
\newblock {\em Physical review letters}, 108(23):235702, 2012.

\bibitem{cates2015motility}
Michael~E Cates and Julien Tailleur.
\newblock Motility-induced phase separation.
\newblock {\em Annu. Rev. Condens. Matter Phys.}, 6(1):219--244, 2015.

\bibitem{palacci2013living}
Jeremie Palacci, Stefano Sacanna, Asher~Preska Steinberg, David~J Pine, and
  Paul~M Chaikin.
\newblock Living crystals of light-activated colloidal surfers.
\newblock {\em Science}, 339(6122):936--940, 2013.

\bibitem{arlt2018painting}
Jochen Arlt, Vincent~A Martinez, Angela Dawson, Teuta Pilizota, and Wilson~CK
  Poon.
\newblock Painting with light-powered bacteria.
\newblock {\em Nature communications}, 9(1):768, 2018.

\bibitem{frangipane2018dynamic}
Giacomo Frangipane, Dario Dell'Arciprete, Serena Petracchini, Claudio Maggi,
  Filippo Saglimbeni, Silvio Bianchi, Gaszton Vizsnyiczai, Maria~Lina
  Bernardini, and Roberto Di~Leonardo.
\newblock Dynamic density shaping of photokinetic e. coli.
\newblock {\em Elife}, 7:e36608, 2018.

\bibitem{schnitzer1993theory}
Mark~J Schnitzer.
\newblock Theory of continuum random walks and application to chemotaxis.
\newblock {\em Physical Review E}, 48(4):2553, 1993.

\bibitem{stenhammar2016light}
Joakim Stenhammar, Raphael Wittkowski, Davide Marenduzzo, and Michael~E Cates.
\newblock Light-induced self-assembly of active rectification devices.
\newblock {\em Science advances}, 2(4):e1501850, 2016.

\bibitem{magiera2015trapping}
Martin~P Magiera and Lothar Brendel.
\newblock Trapping of interacting propelled colloidal particles in
  inhomogeneous media.
\newblock {\em Physical Review E}, 92(1):012304, 2015.

\bibitem{sharma2017brownian}
Abhinav Sharma and Joseph~M Brader.
\newblock Brownian systems with spatially inhomogeneous activity.
\newblock {\em Physical review E}, 96(3):032604, 2017.

\bibitem{grauer2018spontaneous}
Jens Grauer, Hartmut L{\"o}wen, and Liesbeth~MC Janssen.
\newblock Spontaneous membrane formation and self-encapsulation of active rods
  in an inhomogeneous motility field.
\newblock {\em Physical Review E}, 97(2):022608, 2018.

\bibitem{Note3}
This neglects changes in $n$ due to a defect flux $\protect \bm {\nabla }\cdot
  {\protect \bf j}_n$, but it can be accounted for easily.

\bibitem{Note4}
A related phenomenon is the spontaneous charging of isotropic tactoids in an
  active nematic \cite {genkin2018spontaneous}.

\bibitem{opathalage2019self}
Achini Opathalage, Michael~M Norton, Michael~PN Juniper, Blake Langeslay, S~Ali
  Aghvami, Seth Fraden, and Zvonimir Dogic.
\newblock Self-organized dynamics and the transition to turbulence of confined
  active nematics.
\newblock {\em Proceedings of the National Academy of Sciences},
  116(11):4788--4797, 2019.

\bibitem{beris1994thermodynamics}
Antony~N Beris and Brian~J Edwards.
\newblock {\em Thermodynamics of flowing systems: with internal
  microstructure}.
\newblock Number~36. Oxford University Press on Demand, 1994.

\bibitem{mishra2010dynamic}
Shradha Mishra, R~Aditi Simha, and Sriram Ramaswamy.
\newblock A dynamic renormalization group study of active nematics.
\newblock {\em Journal of Statistical Mechanics: Theory and Experiment},
  2010(02):P02003, 2010.

\bibitem{shankar2018low}
Suraj Shankar, Sriram Ramaswamy, and M~Cristina Marchetti.
\newblock Low-noise phase of a two-dimensional active nematic system.
\newblock {\em Physical Review E}, 97(1):012707, 2018.

\bibitem{vromans2016orientational}
Arthur~J Vromans and Luca Giomi.
\newblock Orientational properties of nematic disclinations.
\newblock {\em Soft matter}, 12(30):6490--6495, 2016.

\bibitem{tang2017orientation}
Xingzhou Tang and Jonathan~V Selinger.
\newblock Orientation of topological defects in 2d nematic liquid crystals.
\newblock {\em Soft Matter}, 13(32):5481--5490, 2017.

\bibitem{genkin2018spontaneous}
Mikhail~M Genkin, Andrey Sokolov, and Igor~S Aranson.
\newblock Spontaneous topological charging of tactoids in a living nematic.
\newblock {\em New Journal of Physics}, 20(4):043027, 2018.

\end{thebibliography}

\end{document}